# METAL-INSULATOR-METAL TRANSITIONS, SUPERCONDUCTIVITY AND MAGNETISM IN GRAPHITE

Y. Kopelevich[*], P. Esquinazi[**], J. H. S. Torres[*], R. R. da Silva[*], H. Kempa[**], F. Mrowka[**], and R. Ocaña[**]

[*]Instituto de Física "Gleb Wataghin", Universidade Estadual de Campinas, Unicamp 13083-970, Campinas, São Paulo, Brasil

[**]Abteilung Supraleitung und Magnetismus, Institut für Experimentelle Physik II, Universität Leipzig, Linnéstrasse 5, D-04103 Leipzig, Germany

## I. INTRODUCTION

Three allotrope forms of carbon, viz., graphite, fullerenes, and carbon nanotubes (CNT) have recently attracted a broad scientific interest due to the experimental evidence that superconductivity and ferromagnetism in these materials may occur at temperatures as high as or even above room temperature. The first observation of superconductivity in carbon-based materials was in alkali-metal graphite intercalation compounds (GIC) [1], and the highest transition temperature $T_c$ = 5 K was reached for $C_2Na$ [2]. The observation of superconductivity in alkali-doped $C_{60}$ (buckminsterfullerenes) at 18 K in $K_3C_{60}$ [3] and at 33 K in $Cs_xRb_yC_{60}$ [4] triggered a substantial amount of experimental and theoretical work to understand the properties of these materials in both normal and superconducting states as well as to obtain $C_{60}$-based and/or related materials with higher $T_c$. The apparent local superconductivity that may occur within isolated "grains" in highly oriented pyrolitic graphite (HOPG) samples below and above room temperature has been reported in Refs. [5, 6]. More recent studies demonstrated that superconductivity at temperatures to 37 K can be obtained in sulfur-doped graphite [7-9]. Superconducting transition temperatures of 52 K in $C_{60}$ [10] and $T_c$ = 117 K in a lattice-expanded $C_{60}/CHBr_3$ [11] were reached by means of gate-induced hole doping in a field-effect transistor geometry. The occurrence of superconductivity in carbon nanotubes at 0.55 K has been demonstrated in Ref. [12]. The results reported in Refs. [13-15] suggest that superconducting correlations in CNT can take place at much higher temperatures, viz. at 20 K [13] and apparently at room temperature [14,15].

Besides, a considerable attention to graphite and related materials has been triggered by the discovery of superconductivity at 39 K in $MgB_2$ [16], a material similar to graphite both electronically and crystallographically.

On the other hand, carbon-based materials are of importance from both scientific and technological points of view due to a ferromagnetic behavior which these materials demonstrate at normal conditions. For instance, the ferromagnetic behavior at room temperature has been reported for pyrolitic carbon [17], oxidized $C_{60}$ [18], HOPG [6], polymerized rhombohedral (Rh) $C_{60}$ [19], and $C_{60}H_{24}$ [20]. The results of Ref. [19] provide evidence that the ferromagnetism in Rh-$C_{60}$ develops approaching the Rh-$C_{60}$ – graphitized fullerene boundary on the pressure-temperature phase diagram, which may suggest a common origin for the ferromagnetism in Rh-$C_{60}$ and graphite. Furthermore, an interplay between ferromagnetism and superconductivity has been demonstrated for both HOPG [6] and graphite-sulfur composites [9] which is of a fundamental interest [21- 28].

There are several review articles where superconducting properties of alkali-doped fullerenes (see, e. g., [29, 30]) and graphite [31] have been considered in details. In this Chapter we focus our attention on recent experimental work related mainly to the transport and magnetic properties of pure and sulfur-doped graphite. We also comment on modern theoretical models which provide us with a possible understanding of the magnetic-field-induced metal-insulator transiton (MIT) as well as superconducting and ferromagnetic instabilities observed in these materials.

This review is further divided in the following sections:

II     Samples and experimental details
III.    Magnetic-field-induced metal-insulator-type transition
III.1.   The main characteristics of the temperature dependence of the resistance in transverse fields
III.2.   The MIT driven mainly by the transverse to the basal planes field
III.3.   Two-dimensional scaling
III.4.   Thermally activated behavior of the basal-plane resistance at low fields
III.5.   Speculating on the origin of the MIT
IV.    Landau-level-quantization-driven insulator-metal and metal-insulator transitions
V.     The interlayer magnetoresistance of graphite: coherent and incoherent transport
VI.    Magneto-thermal conductivity in HOPG near the metal-insulator transition
VII.   Evidence for superconducting and ferromagnetic correlations: magnetization studies
VII.1.  Superconducting and ferromagnetic instabilities in graphite
VII.2.  Local superconductivity in graphite-sulfur composites
VII.3.  Possible origin of ferromagnetism in graphite

II. SAMPLES AND EXPERIMENTAL DETAILS

Both HOPG and single crystalline Kish graphite samples have been studied. The HOPG samples were obtained from the Research Institute "Graphite" (Moscow), Union Carbide Co. and Advanced Ceramics (AC) from USA. X-ray diffraction ($\Theta$-$2\Theta$) measurements of HOPG samples revealed a characteristic hexagonal graphite structure with no signature of a rhombohedral phase [32]. The high degree of crystallites orientation along the hexagonal c-axis was confirmed for HOPG samples from x-ray rocking curves (FWHM = 0.4…1.4$^o$). Both magnetization and resistance measurements were performed for H ∥ c-axis and H ⊥ c-axis applied magnetic field configurations. In some cases, the measurements as a function of angle $\gamma$ between H and the sample basal planes were performed. The magnetization M(T, H, $\gamma$) was measured in fields up to 7 T and temperatures between 2 and 800 K by means of the SQUID magnetometers MPMS5 and MPMS7 (Quantum Design). Low-frequency (f = 1 Hz) and dc both longitudinal measurements of $R_{xx}$(H, T, $\gamma$) and Hall $R_{xy}$(H, T, $\gamma$) resistances were performed in fields up to 9 T in the same temperature interval using Quantum Design PPMS-9T, Janis-9T magnet systems, Oxford 9 T-magnet systems with home-made inserts. Some measurements have been done in the mK region using a dilution refrigerator with a 8 T superconducting solenoid.

Graphite-sulfur (C-S) composites were prepared by mixture of the graphite powder consisting of ~ 8 $\mu$m size particles [the impurity content in ppm: Fe (32), Mo (< 1 ), Cr (1.1), Cu (1.5)] and the sulfur powder (99.998 %; Aldrich Chemical Company, Inc.) in a ratio C:S = 1:1. The mixture was pressed into pellets, held under Ar atmosphere at 650 K for one hour and subsequently annealed at 400 K for 10 hours before cooling to room temperature. The final sulfur contents in the composite was 23 wt %. X-ray ($\Theta$ - $2\Theta$ geometry) measurements revealed a small decrease in the c-axis parameter of the hexagonal graphite from c = 6.721 Å in the pristine graphite powder to c = 6.709 Å in the composite sample, and no changes in the lattice parameters of the orthorhombic sulfur (a = 10.45 Å, b = 12.84 Å, c = 24.46 Å). No impurity or additional phases were found.

III. MAGNETIC – FIELD - INDUCED METAL - INSULATOR - TYPE TRANSITION

III.1. The main characteristics of the temperature dependence of the resistance for transverse fields

The transition from metallic- ($dR_b/dT > 0$) to insulator-like ($dR_b/dT < 0$) behavior of the basal-plane resistance $R_b$(T, H) driven by a magnetic field applied perpendicular to the graphene planes has first been reported in Ref. [5] for HOPG-1 sample having the basal-plane resistivity $\rho_b$(T = 300 K, H = 0) ≈ 45 $\mu\Omega$cm, and the room temperature resistivity ratio $\rho_c/\rho_b$ ≈ 8.56 x 10$^3$, where $\rho_c$ is the resistivity along the c-axis direction.

The field-induced suppression of the metallic state has been attributed to the suppression of superconducting correlations in graphite [5]. As can be seen from Fig. 1, at low enough magnetic fields (H < 800 Oe), the $R_b(T)$ measured for this sample demonstrates a maximum at a field-dependent temperature $T_{max}(H)$. At not too high magnetic fields but higher than 800 Oe, an insulating ($dR_b/dT < 0$) resistance behavior was observed in the explored temperature interval 2 K ≤ T ≤ 300 K.

A similar metal-insulator-type transition (MIT) induced by the applied magnetic field has been observed for all studied graphite samples. Figures 2, 3, and 4 illustrate the MIT occurrence in another HOPG sample from the "Graphite" Institute, labeled as HOPG-3 (Fig. 2), HOPG sample from the Union Carbide Co., labeled as HOPG-UC (Fig. 3), and a single crystalline Kish graphite sample, labeled as KISH-1 (Fig. 4).

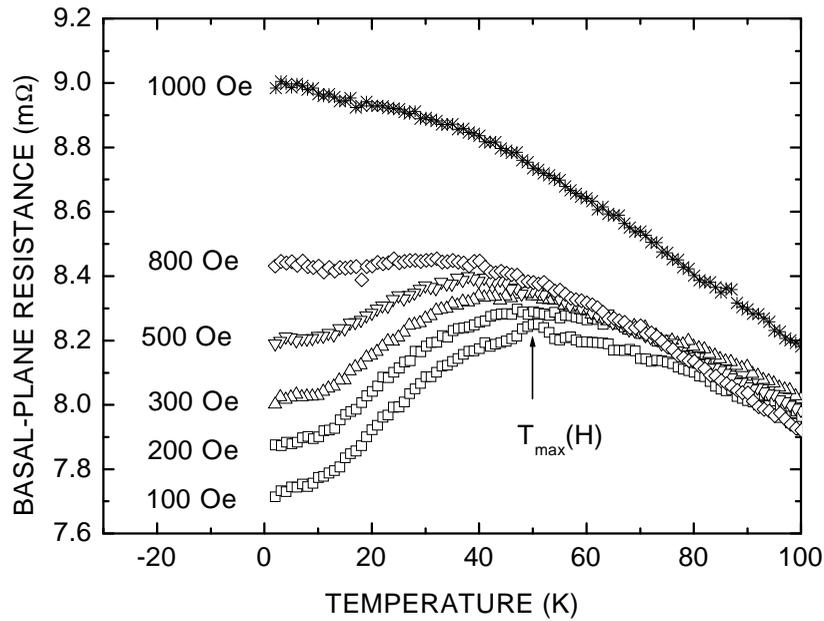

Fig. 1. Basal-plane resistance $R_b(T, H)$ measured for the sample HOPG-1 at various magnetic fields applied perpendicular to the graphene planes. $T_{max}(H)$ separates a high-temperature insulator-like ($dR_b/dT < 0$) from a low-temperature metallic-like ($dR_b/dT > 0$) resistance behavior.

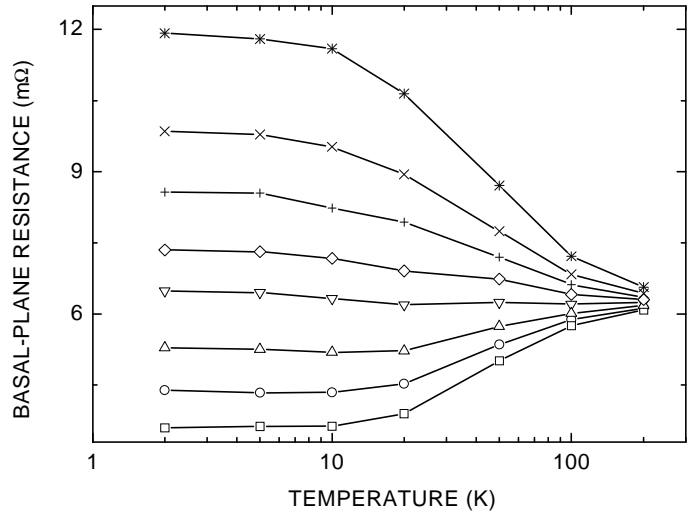

Fig. 2. Basal-plane resistance $R_b(T, H)$ measured for the sample HOPG-3 at various magnetic fields (from bottom to the top: H = 500, 700, 900, 1140, 1300, 1500, 1700, and 2000 Oe) applied perpendicular to the graphene planes; $\rho_b(T = 300 K, H = 0) \approx 5\ \mu\Omega cm$ and $\rho_c/\rho_b \approx 5 \times 10^4$.

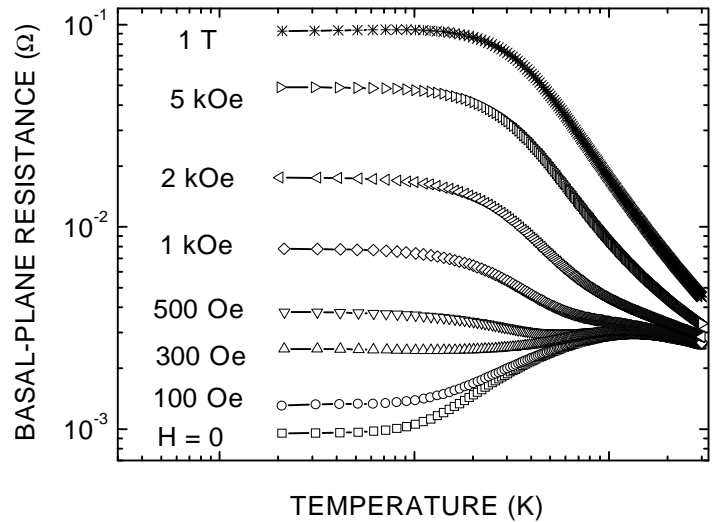

Fig. 3. Basal-plane resistance $R_b(T, H)$ measured for the sample HOPG-UC at various magnetic fields applied perpendicular to the graphene planes; $\rho_b(T = 300 K, H = 0) \approx 3\ \mu\Omega cm$ and $\rho_c/\rho_b \approx 5 \times 10^4$.

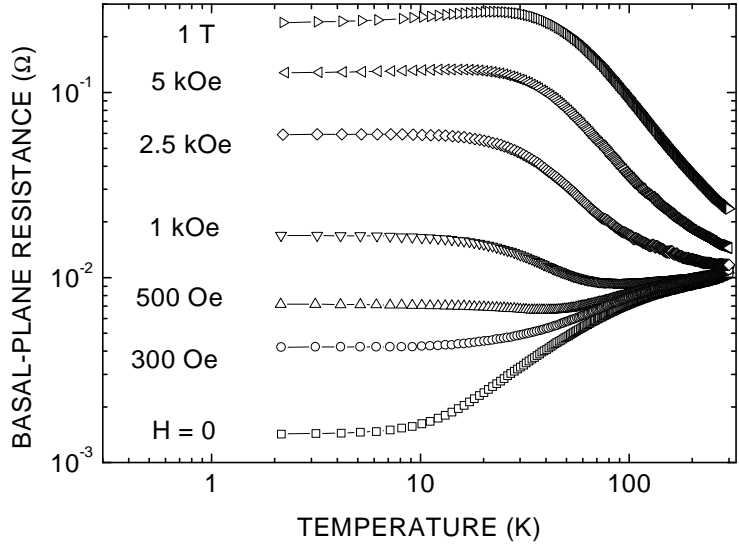

Fig. 4. Basal-plane resistance $R_b(T, H)$ measured for the KISH-1 single crystal at various magnetic fields applied perpendicular to the graphene planes; $\rho_b(T = 300$ K, $H = 0) \approx 5$ μΩcm and $\rho_c/\rho_b \approx 10^2$.

From Figs. 2 – 4 we conclude that the graphite samples with a low zero-field in-plane resistivity demonstrate:
(1) a metallic behavior in a broad temperature range,
(2) a giant magnetoresistance at low temperatures,
(3) a magnetic-field-induced MIT, and
(4) the development of a minimum in $R_b(T)$ which occurs increasing field.

Figure 5(a) and Fig. 5(b) provide a detailed view of the resistance minimum appearance at the temperature $T_{min}(H)$, which is an increasing function of field, for HOPG-UC and KISH-1 samples. Figure 6 presents the $T_{min}(H)$ obtained for these two samples. The data of Fig. 5(a, b) and Fig. 6 provide also evidence for the existence of a threshold field $H_c$ below which the metallic state of graphite is preserved. The dotted and solid lines in Fig. 6 demonstrate that above and in the vicinity of $H_c$, $T_{min}(H)$ obtained for the KISH-1 sample can be well described by the equations

$$T_{min}(H) = A\,(H - H_c)^{1/2} \qquad (1)$$

and

$$T_{min}(H) = B\,[1 - (H_c/H)^2]\,H^{1/2}, \qquad (2)$$

where A, B and $H_c$ being fitting parameters. $T_{min}(H)$ obtained for HOPG-UC will be analyzed below. The Eqs. (1) and (2) are predicted by theories [35, 36] which assume that $T_{min}(H)$ corresponds to the metal-insulator transition temperature $T_c(H)$, as well as that the MIT in graphite is a manifestation of the so-called "magnetic catalysis" (MC) phenomenon known in the relativistic theories of the (2 + 1) – dimensional Dirac fermions.

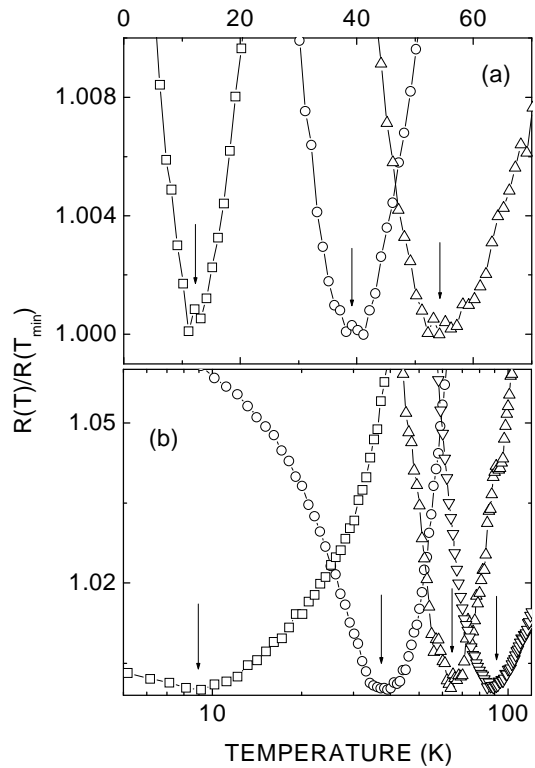

Fig. 5. (a) Reduced basal-plane resistance $R(T)/R(T_{min})$ measured for HOPG-UC with H = 300 Oe (□), H = 400 Oe (o), H = 500 Oe (Δ); (b) $R(T)/R(T_{min})$ measured for Kish graphite single crystal with H = 400 Oe (□), H = 500 Oe (o), H = 750 Oe (Δ), and H = 1 kOe (∇). Arrows (a, b) indicate $T_{min}(H)$, the field-dependent temperature which separates insulating-like (T < $T_{min}$) and metallic-like (T > $T_{min}$) resistance behavior. Magnetic field is always applied parallel to the sample c-axis.

More specifically, according to the theory [35, 36], the magnetic field applied perpendicular to the graphene planes opens an insulating gap in the spectrum of Dirac fermions, associated with an electron-hole pairing. Such an approach is appealing because the characteristic feature of the band structure of a single graphene layer is that there are two isolated points in the first Brillouin zone where the band dispersion is linear $E(\mathbf{k}) = \hbar v |\mathbf{k}|$ ($v = v_F \sim 10^6$ m/s is the Fermi velocity), so that the electronic states can be described in terms of Dirac equations in two dimensions (2D) [35 - 38].

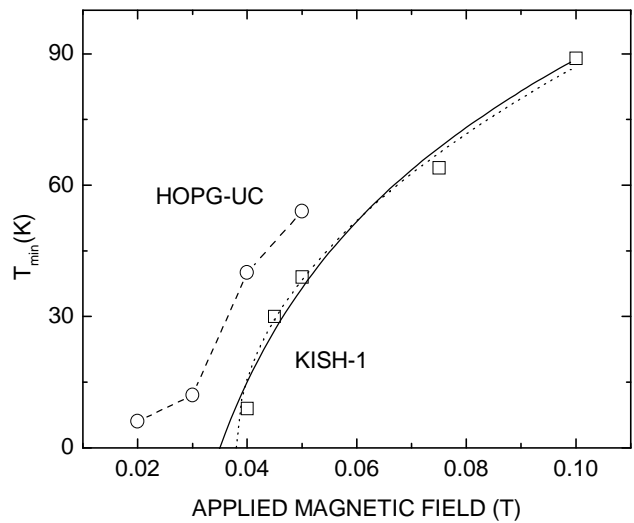

Fig. 6. $T_{min}(H)$ obtained for HOPG-UC (o) and KISH-1 (□) samples; dotted line is obtained from Eq. (1) with the fitting parameters $A = 350$ K/T$^{1/2}$, $\mu_o H_c = 0.038$ T, and the solid line is obtained from Eq. (2) with the fitting parameters $B = 320$ K/T$^{1/2}$, $H_c = 0.035$ T. The magnetic field is applied perpendicular to the graphite basal planes.

III.2. The MIT driven mainly by the transverse to the basal planes field

The results presented below provide evidence that mainly the magnetic field component perpendicular to basal graphite planes drives the MIT; Figs. 7 - 10 illustrate this fact. Figures 7 and 8 present $R_b(T,H)$ measured for the HOPG-UC sample with magnetic field applied along the planes within an experimental misalignment of ~ 2 degrees, demonstrating in particular that the insulating state (the minimum in $R_b(T,H)$) takes place at much higher fields as compared to that measured in H ∥ c-axis geometry, see Fig. 3 and Fig. 5 (a).

In Fig. 9 we have plotted $T_{min}(H)$ and $T_{max}(H)$ [$T_{max}(H)$ is defined similar to that shown in Fig. 1 for HOPG-1 sample] obtained for the HOPG-UC sample in both H || c-axis and H || planes geometry. Figure 10 presents the same data re-plotted as a function of the field component perpendicular to graphene planes, taking a misalignment angle between the applied field and basal planes $\gamma = 1.7°$ in "H || planes" geometry, indicating that the data appear to scale with $H\sin\gamma$. Recent high-resolution angle measurements show, however, that not only the extrinsic but an intrinsic crystal misalignment, characterized by the rocking curve of the sample, is important to understand the observations at parallel fields. These measurements show also that the data do not rigorously follow the simple scaling shown in Fig. 10. It appears that the best approach to prove that only the transverse field component is important for the MIT is to show that the "critical" field at which $T_{max}(H)$ and $T_{min}(H)$ coincide scales with that field component. Whatever the method used, the experimental data indicate that mainly orbital (in-plane) effects drive the MIT in graphite, which would support the MC-based explanation of the effect [35, 36].

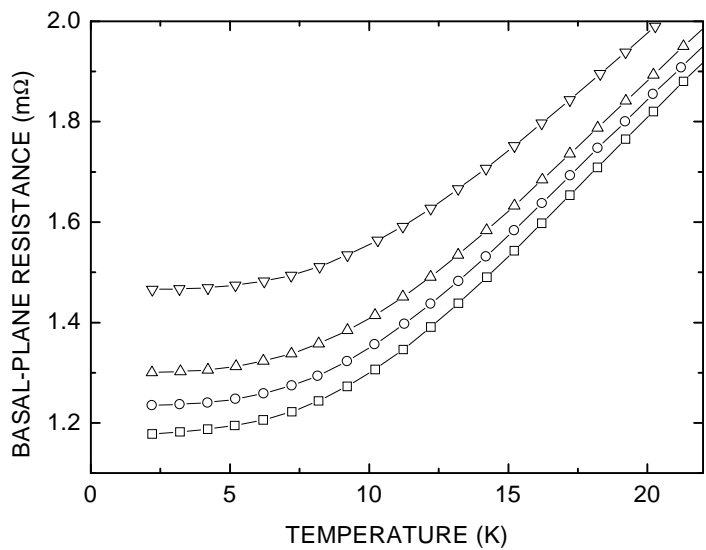

Fig. 7. Low-temperature portions of the basal-plane resistance $R_b(T, H)$ measured for the sample HOPG-UC at H = 100 Oe (□), H = 500 Oe (o), H = 1 kOe (Δ), and H = 2 kOe (∇) applied along the graphene planes.

Furthermore, Fig. 10 indicates that $T_{min}(H)$ measured for both H || c and H ⊥ c configurations deviates from the behavior described by Eq.(1) or Eq.(2). Such a deviation has also been observed for the KISH-2 single crystal which is from the same batch as the KISH-1 sample. Figure 11 presents $T_{min}(H)$ measured for both HOPG-UC and KISH-2 samples. As can be seen from that plot, $T_{min}(H)$ can be described by Eqs. (1) and (2) only in the vicinity of a critical field $H_c$.

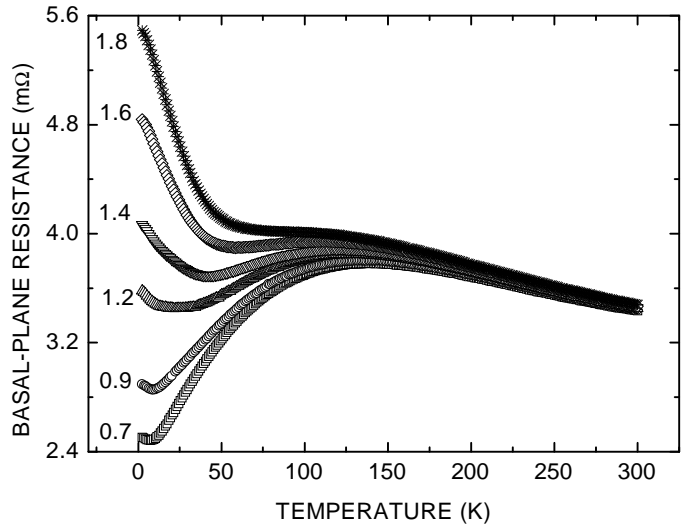

Fig. 8. Basal-plane resistance $R_b(T, H)$ measured for the sample HOPG-UC at various magnetic fields (in Tesla) applied along the graphene planes.

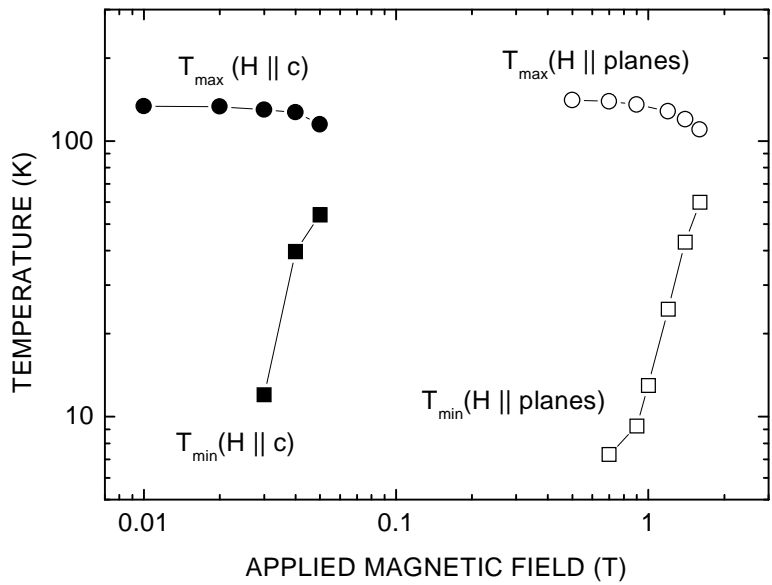

Fig. 9. $T_{min}(H)$ and $T_{max}(H)$ measured for HOPG-UC sample with magnetic field applied either parallel (H || c) or perpendicular (H || planes) to the sample c-axis.

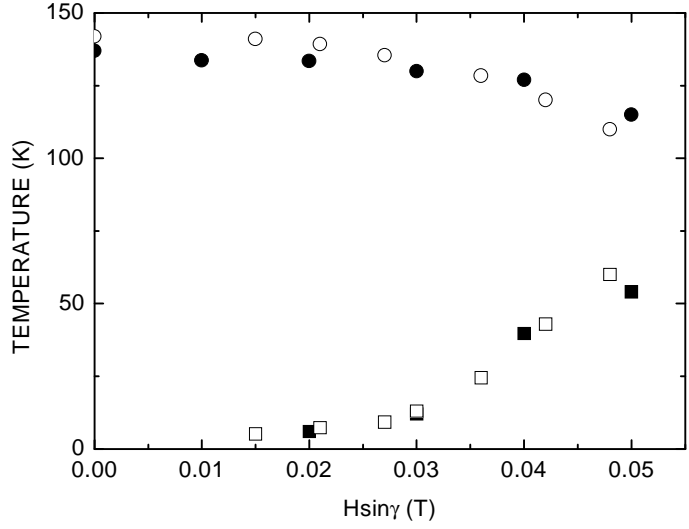

Fig. 10. The data of Fig. 9 re-plotted as a function of the field component perpendicular to the graphite basal planes $H_\perp = H\sin\gamma$, where $\gamma$ is the angle between applied magnetic field and the planes; $\gamma = 90°$ ($H \parallel c$) and a misalignment angle $\gamma = 1.7°$ is taken in the case of "$H \parallel$ planes" geometry.

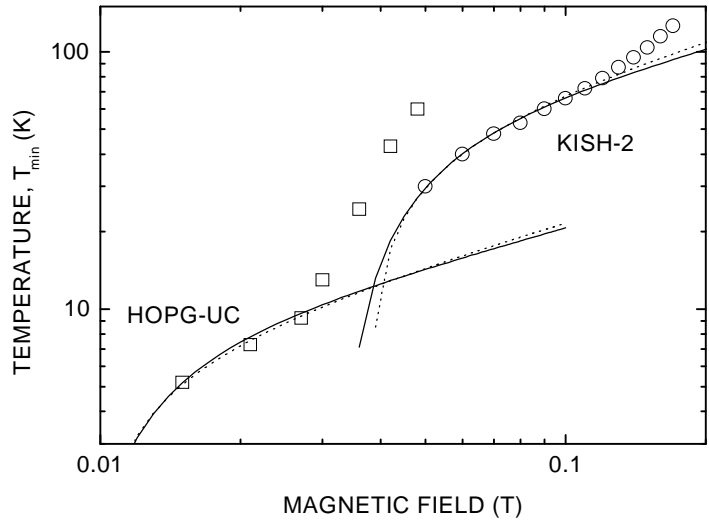

Fig. 11. $T_{min}(H)$ measured for HOPG-UC and KISH-2 samples; for HOPG-UC $T_{min}(H)$ is obtained in $H \parallel$ planes geometry and plotted as a function of the field component parallel to the sample c-axis; the results for KISH-2 sample were obtained with $H \parallel$ c-axis. Dotted lines are obtained from Eq. (1) with the fitting parameters $A = 72$ K/T$^{1/2}$, $\mu_o H_c = 0.01$ T (HOPG-UC) and $A = 270$ K/T$^{1/2}$, $\mu_o H_c = 0.038$ T (KISH-2); solid lines correspond to the Eq. (2) taking $B = 66$ K/T$^{1/2}$, $\mu_o H_c = 0.009$ T (HOPG-UC), and $B = 235$ K/T$^{1/2}$, $\mu_o H_c = 0.033$ T (KISH-2).

III.3. Two-dimensional scaling

Our analysis revealed also [33] that the scaling approach used to characterize both the magnetic-field-induced superconductor-insulator quantum phase transition [39] and the field-driven MIT in 2D electron (hole) systems [40] can be equally well applied to the MIT observed in graphite. According to the scaling theory of the superconductor-insulator transition [39, 41, 42], the resistance in the critical regime of the T = 0 transition is given by the equation:

$$R(\delta, T) = R_{cr} f(|\delta|/T^{1/z\nu}),  \qquad (3)$$

where $R_{cr}$ is the resistance at the transition, $f(|\delta|/T^{1/z\nu})$ a scaling function such that $f(0) = 1$; $z$ and $\nu$ are critical exponents, and $\delta$ the deviation of a variable parameter from its critical value. With $\delta = H - H_{cr}$ we have plotted R vs. $|\delta|/T^{1/\alpha}$ for the HOPG-3 sample in Fig. 12, where $\alpha = 0.65 \pm 0.05$ was extracted from log-log plots of $(dR/dH)|_{H_{cr}}$ vs. $T^{-1}$, and the critical field $H_{cr} = 1.14$ kOe was obtained from experimental data presented in Fig. 13. As can be seen in Fig. 12, the resistance data obtained in the temperature range 50 K - 200 K collapse into two distinct branches, below and above the $H_{cr}$. At T < 20 K, where the resistance $R_b(T)$ saturates, a clear deviation from the scaling takes place, reminiscent of the behavior observed in amorphous Mo-Ge films [43] and described in terms of a 2D Bose-metal [44-49]. In the case of graphite and Si MOSFETS's [40] the minimum in R(T), which develops increasing field, see e. g. Fig. 5, implies that Eq. (3) is applicable only in a restricted temperature range. The analysis performed on various HOPG as well as Kish graphite samples revealed the universality of the above scaling. Figure 14(a) and Fig. 14(b) exemplifies both the field-driven MIT and the scaling observed in the KISH-2 sample.

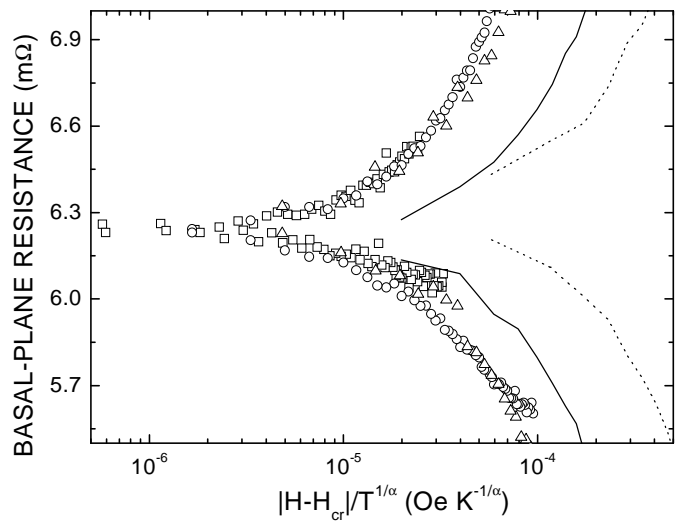

Fig. 12. Basal-plane resistance measured for HOPG-3 at T = 10 (- -), 20 (—), 50 (Δ), 100 (o), and 200 K (□) plotted vs. the scaling variable, where $H_c = 1140$ Oe and $\alpha = 0.65$.

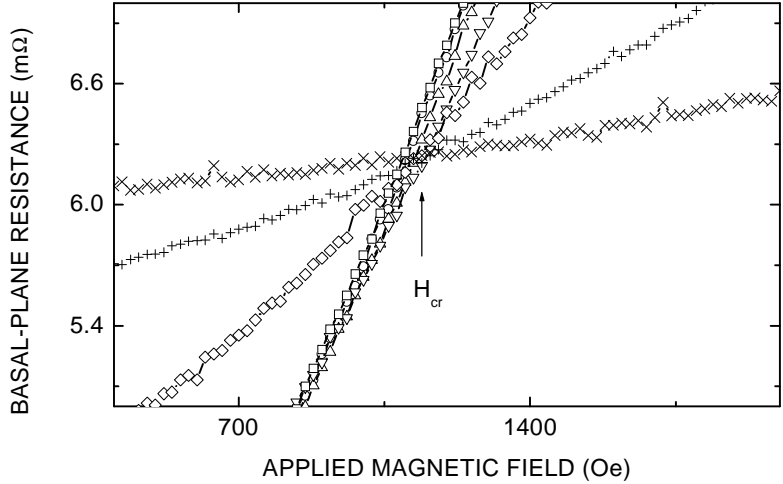

Fig. 13. Magnetoresistance $R_b(H)$ measured in HOPG-3 at T = 2 (□), 5 (o), 10 (Δ), 20 (∇), 50 (◊), 100 (+), and 200 K (x); $H_{cr}$ = 1140 Oe is the "critical" field.

For all graphite samples the obtained value of the exponent α = 0.65 ± 0.05 coincides with that found in the scaling analysis of both field-driven MIT in Si MOSFET (α = 0.6 ± 0.1) [50] and the superconductor-insulator transition in ultrathin bismuth films (α = 0.7 ± 0.2) [51].

III.4. Thermally activated behavior of the basal-plane resistance at low fields

There is another important similarity between the resistance $R_b(T,H)$ behavior in graphite and that in other 2D systems. Namely, we found that $R_b(T,H)$ at zero or small enough applied fields and $T < T_{max}(H)$ can be described by the equation (see Fig. 15):

$$R(T) = R_0(H) + R_1(H)\exp(-E_a(H)/k_BT), \quad (4)$$

where $R_0(H)$, $R_1(H)$, and $E_a$ are fitting parameters. A better fit can be obtained adding to Eq. (4) a term of the form $R_2(H)\exp(-E_b(H)/k_BT)$. For simplicity we restrict here the discussion considering only the term with the activation energy $E_a(H)$. Equation (4) implies a thermally-activated metallic resistance behavior in graphite, also observed in e.g. Si MOSFET [52], p-type GaAs [53], and p-type SiGe [54].

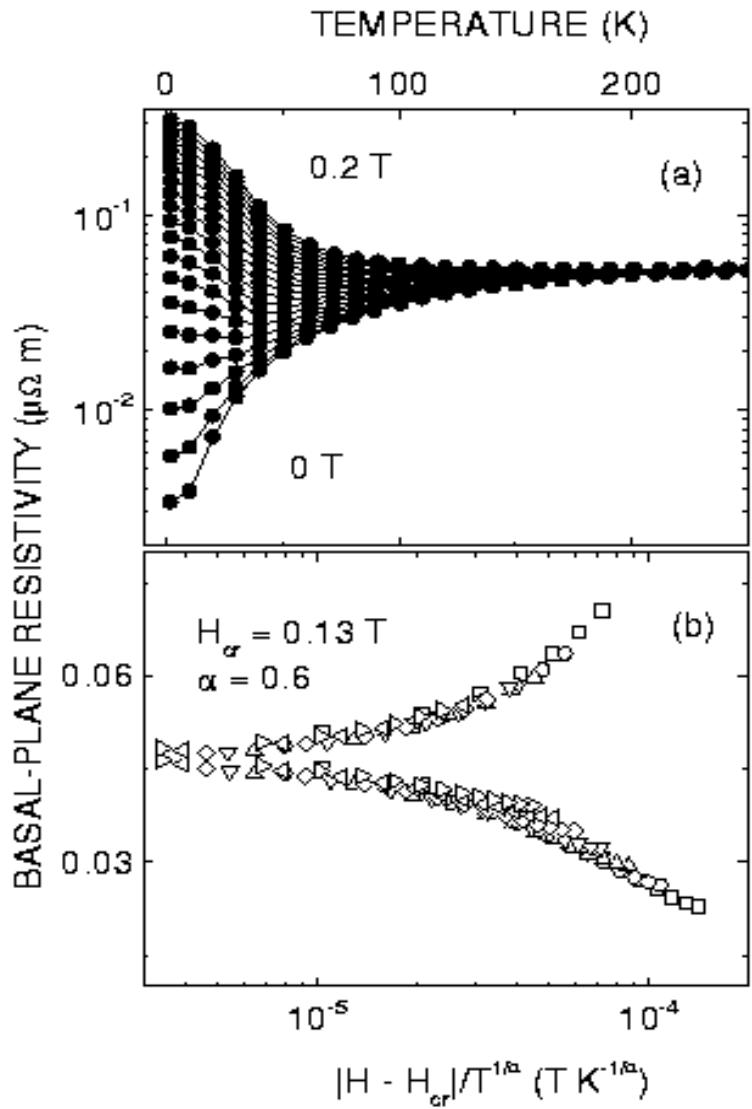

Fig. 14. (a) Basal-plane resistivity $\rho_b(T,H)$ measured in KISH-2 single crystal (the field step between two neighboring curves is 0.01 T); (b) scaling behavior of the resistivity (see text) in the temperature range between 50 K and 130 K.

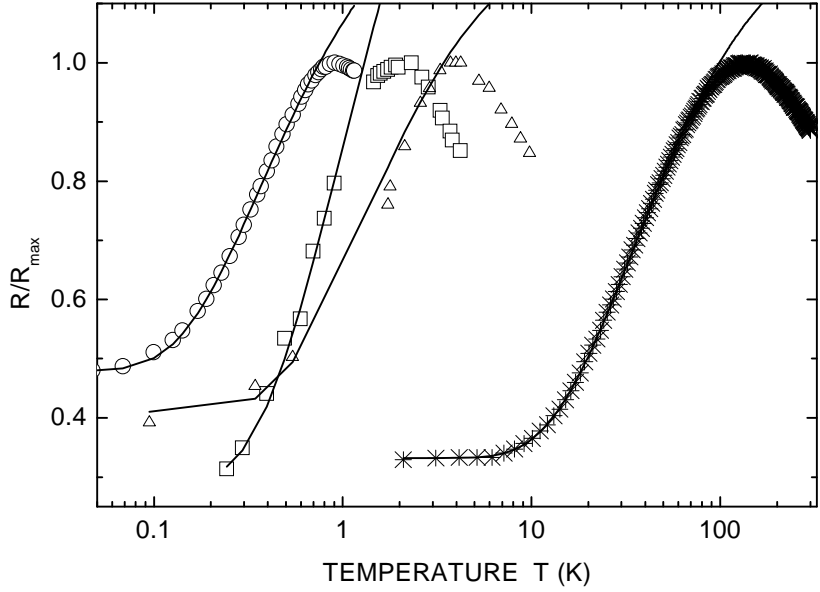

Fig. 15. Normalized resistance $R(T)/R_{max}(T=T_{max})$ obtained at no applied magnetic field for HOPG-UC (*), Si MOSFET (□) [52], GaAs (o) [53], and SiGe (Δ) [54]. Solid lines correspond to Eq. (4) normalized by $R_{max}$ with $R_0$ = 0.97 mΩ (HOPG-UC), 12.955 kΩ/□ (Si MOSFET), 1.59 kΩ/□ (GaAs), 4.43 kΩ/□ (SiGe); $R_1(0)$ = 2.75 mΩ (HOPG-UC), 6.7 kΩ/□ (Si MOSFET), 2.8 kΩ/□ (GaAs), 9.18 43 kΩ/□ (SiGe), and $E_a$ = 34 K (HOPG-UC), 0.97 K (Si MOSFET), 0.37 K (GaAs), 1.25 K (SiGe).

Estimating the Fermi temperature $T_F = \pi\hbar^2 n_{2D}/2k_B m^*$ = 7.5 K ($n_{2D}$ is the 2D carrier density and $m^*$ the effective mass) for SiGe (p = 1.2 x $10^{11}$ cm$^{-2}$, $m^* = 0.22m_0$), $T_F$ = 1.1 K for GaAs (p = 0.3 x $10^{11}$ cm$^{-2}$, $m^* = 0.38m_0$), and $T_F$ = 6 K for Si MOSFET (n = 0.88 x $10^{11}$ cm$^{-2}$, $m^* = 0.2m_0$), and taking $T_F$ = 240 K for graphite [55], one can see from Fig. 15 that the exponential metallic behavior in all these systems takes place at T < $T_F$. The ratio $E_a/E_F$ = 0.14 found for HOPG-UC practically coincides with that obtained for Si MOSFET (~ 0.16) and SiGe (~ 0.17); these values differ by a factor ~ 2 from $E_a/E_F$ = 0.34 estimated for GaAs. On the other hand, whereas $E_a$ decreases by increasing the parallel field in Si MOSFET [56], and is field independent in GaAs [57], $E_a(H)$ in graphite increases with field. Figure 16 illustrates this fact where $R_b(T, H)$ for HOPG-UC sample grows exponentially with temperature at zero and low enough applied magnetic fields. The inset in Fig. 16 presents the activation energy $E_a(H)$. Because of the minimum in $R_b(T)$, occurring at higher fields, no unambiguous conclusion regarding $E_a(H)$ at $H_\perp \geq 200$ Oe can be drawn.

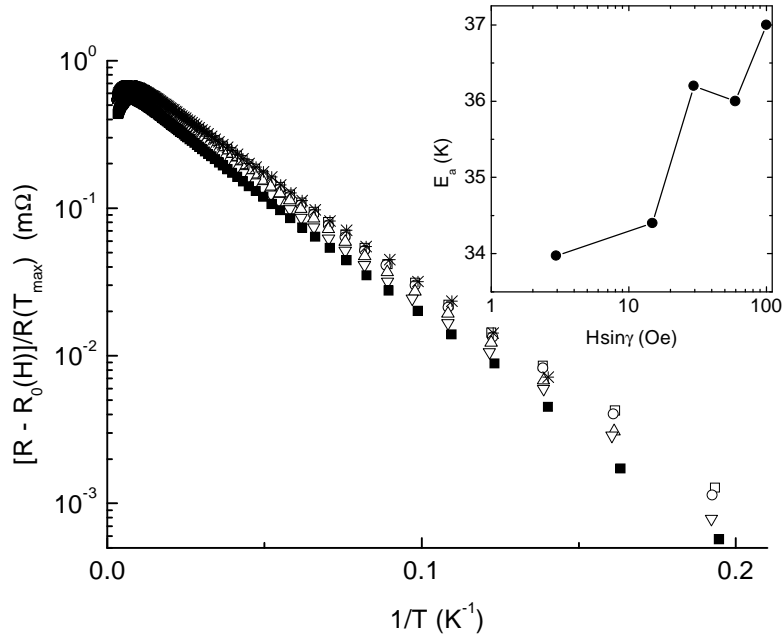

Fig. 16. Reduced basal-plane resistance vs. 1/T measured for HOPG-UC sample at H = 0 (*), H = 100 Oe (□), H = 500 Oe (o), H = 1 kOe (Δ), H = 2 kOe (∇) applied along the planes (the misalignment angle $\gamma = 1.7°$), and H = 100 Oe (■) applied parallel to the c-axis ($\gamma = 90°$), demonstrating the thermally-activated metallic resistance behavior in graphite. Inset shows the activation energy $E_a$ (in K) plotted against $H_\perp = H\sin\gamma$.

III.5. Speculating on the origin of the MIT

The mechanism behind the zero-field gap $E_a$ and its field dependence has not been clarified yet. We stress, however, that if the electronic spectrum of graphite is gapped at zero applied magnetic field, as experiments suggest, the MC-based interpretation of the MIT in graphite [35, 36] might need to be revised.

According to Ref. [50], the metallic state in 2D systems such as Si MOSFET's is a Bose-metal state with Cooper pairs lacking a phase coherence. The striking similarity between graphite and other 2D systems demonstrating the superconductor-insulator-type transition, may indicate the occurrence of a Bose-metal state in graphite as well. Magnetization measurements [6-9] revealed the existence of superconducting "grains" in graphite, indeed. The exponential increase with temperature of the resistance and its saturation at low temperatures have been observed in various systems with Cooper pairs such as MoGe films [43] and artificially grown Josephson-junction arrays (JJA) [58].

Among the possible mechanisms for the Cooper pair formation in quasi-2D systems, we would like to mention the superconductivity of anyons [59, 60] (particles with a fractional statistics) [61, 62]. Because of the expected resonating-valence-bond (RVB) ground state of graphite on one hand [63, 64], and the identification of the RVB state with anyon superconductivity and the fractional quantum Hall effect (FQHE) on the other hand [65 - 67], the proposal by Zhang and Rice [59, 60] may have a particular relevance to the case of a quasi-2D graphite. Ren and Zhang [68] have shown that the anyon superconductivity should occur in 2D electron gas with the Coulomb coupling constant $r_s > 4$.

For the quasi-two-dimensional hexagonal graphite we have

$$r_s = ((\pi n_{2D})^{1/2} a_B^*)^{-1} , \qquad (5)$$

where $n_{2D} = n_{3D}d$ is the 2D-carrier density, $d = 3.35$ Å is the interplane distance, $a_B^* = \varepsilon \hbar^2/m^* e^2$ is the effective Bohr radius, and $\varepsilon = 2.8$ [55] is the dielectric constant. The 3D-density of the majority carriers $n_{3D} \sim 2 \times 10^{18}$ cm$^{-3}$ and their effective mass $m^* \sim 0.05 m_0$ ($m_0$ is the free-electron mass). For the minority carriers we have $n_{3D} \sim 6 \times 10^{16}$ cm$^{-3}$ and $m^* \sim 0.004 m_0$ [5]. Substituting the above parameters in Eq. (5) one gets $r_s \sim 5$ and $r_s \sim 10$ for the majority and minority carriers, respectively. These numbers indicate that superconductivity may occur in this system [68]. We note also that orbital ferromagnetism can emerge in the anyon system [61].

The existence in graphite of both electrons and holes with a large difference in the magnitude of their effective mass $m^*$ (compare $m^*$ for majority and minority carriers) may lead to the "metallic excitonium" phase [69] where superconductivity at high temperatures is possible [70].

An extrinsic origin of high-temperature superconductivity related to a topological disorder in graphene layers has been proposed in Ref. [71], where it has been shown that topological disorder enhances the density of states at the Fermi level leading, in the presence of sufficiently strong repulsive electron-electron interaction, to a p-wave superconductivity mediated by ferromagnetic spin fluctuations. This prediction is particularly interesting because the experiment provides evidence for the interplay between superconducting and ferromagnetic behavior in both pure graphite [6] and graphite-sulfur composites [9].

## IV. LANDAU-LEVEL-QUANTIZATION-DRIVEN INSULATOR-METAL AND METAL- INSULATOR TRANSITIONS

While the low-magnetic-field-driven MIT in graphite is an intriguing phenomenon by itself, perhaps not less surprising is the observation of a re-entrant metallic ($dR_b/dT > 0$) state in the quantum limit, viz. when carriers occupy only the lowest Landau levels [5]. Aiming to explore the field-induced re-entrant metallic state, we have performed a high-field (H ∥ c-axis) magnetoresistance measurements on several HOPG and Kish graphite samples. The obtained results can be summarized as follows:

(1) the occurrence of magnetic-field-induced metal-insulator and insulator-metal transitions (or crossovers) is a generic property of graphite;

(2) a reentrant metallic state takes place in the Landau level quantization regime for majority carriers, below a temperature $T_{max}(H)$ which is an oscillating or increasing function of field as found for HOPG and Kish graphite samples, respectively;

(3) in HOPG, but not in Kish graphite samples, a sequence of field-driven metal-insulator-metal transitions is found which resembles the quantum Hall effect (QHE) – insulator transitions observed in various (quasi-) 2D electron (hole) systems.

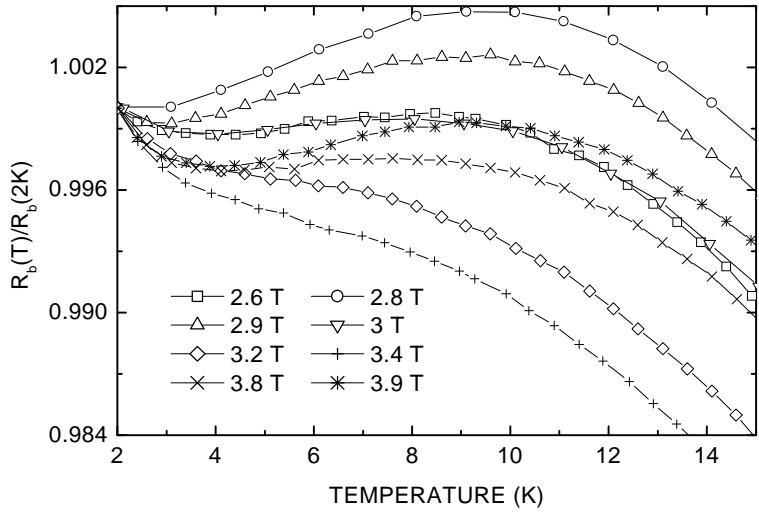

Fig. 17. Normalized basal-plane resistance measured for HOPG-1 sample [5] demonstrating a re-entrant metallic ($dR_b/dT > 0$) resistance behavior at certain applied magnetic fields and temperatures.

Figures 17 –20 present basal-plane resistance $R_b(T,H)$ measured for HOPG-1, HOPG-3, and KISH-1 graphite samples which illustrate the re-appearance of a metallic ($dR_b/dT > 0$) behavior increasing field and at $T < T_{max}(H)$. The $T_{max}(H)$ obtained for four samples (HOPG-1, HOPG-3, HOPG-UC, and KISH-1) are given in Fig. 21. The fact that the reentrant metallic state occurs in the regime of pronounced Landau level quantization can be seen in Figs. 22 - 26 (see also Ref. [5]) where longitudinal [$R_{xx}(H) \equiv R_b(H)$] and Hall [$R_{xy}(H)$] resistance oscillations, usually associated with the Shubnikov - de Haas (SdH) [72 – 75] effect, are shown for HOPG-3 (Figs. 22 - 25) and KISH-1 (Figs. 26, 27) samples. The Hall resistance was measured in van der Pauw geometry using both the cyclic transposition of current and voltage leads [76 - 80] at fixed applied field polarity and a magnetic field reversal; no difference in $R_{xy}(H,T)$ obtained with the two methods was found.

The cyclotron to thermal energy ratio $E_c/E_T = \hbar eH/m^*k_B T$ defines whether the classical magnetotransport approach is appropriate or Landau quantization has to be taken into account. The dotted line in Fig. 21 obtained from the equation

$$T_{cr} = \hbar eH/m^*k_B \qquad (6)$$

separates the classical ($T > T_{cr}$) and quantum ($T < T_{cr}$) regimes illustrating that the re-entrant metallic state(s) take(s) place in the quantum regime.

The puzzling metallic state at high fields persists in clean Kish graphite samples at least to ~ 500 mK [75]. In HOPG samples we have observed an upturn of the resistance $R_b(T, H)$ below the field-independent temperature of ~ 4 K and ~ 1 K for HOPG-1 (Figs. 17, 18) and HOPG-3 (Fig. 19) samples, respectively. Below the minimum, $R_b(T,H)$ decreases logarithmically with temperature, as exemplified in Fig. 19 for $\mu_o H = 2$ T and $\mu_o H = 8$ T. We stress that no such a minimum in the temperature dependence of the resistance was observed at zero applied field down to 70 mK.

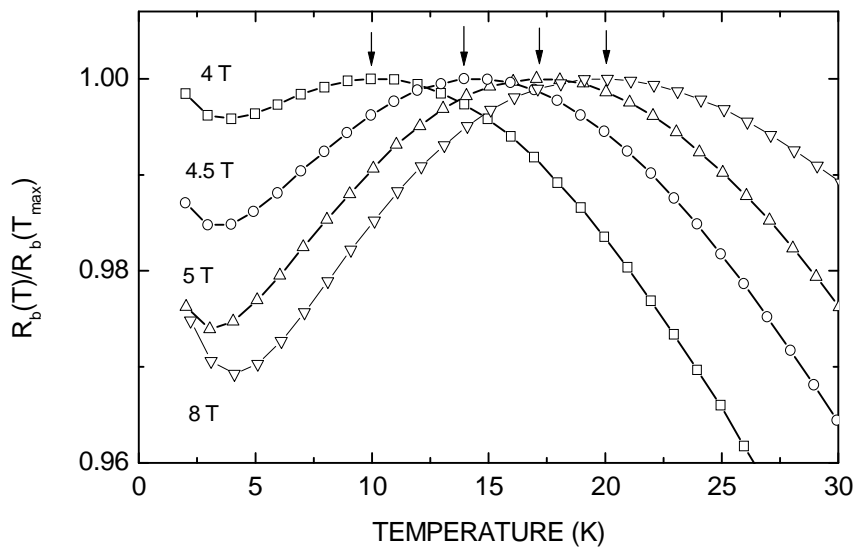

Fig. 18. Normalized basal-plane resistance measured for sample HOPG-1 [5] in the quantum limit; arrows indicate $T_{max}(H)$.

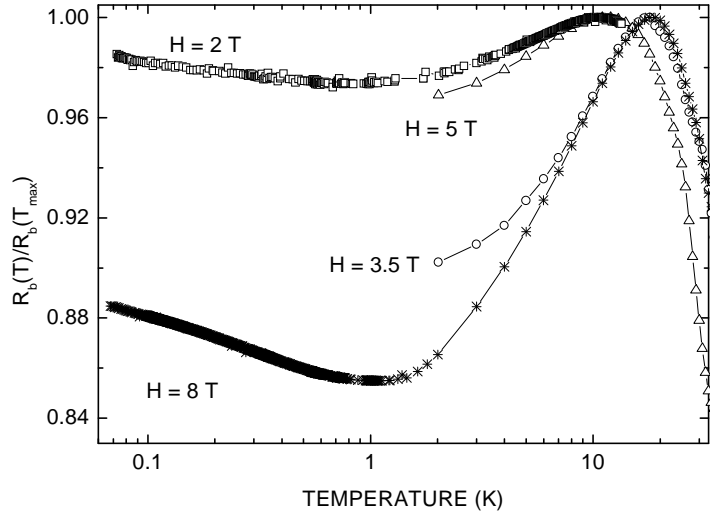

Fig.19. Normalized basal-plane resistance measured for sample HOPG-3 in the high-field regime where the re-entrant metallic state emerges at $T < T_{max}(H)$; $T_{max}(H)$ is defined as in Fig. 18.

It can be seen in Fig. 21 that $T_{max}(H)$ is an oscillating function of field for HOPG samples whereas it increases monotonically with field in Kish graphite (see also Fig. 20).

Both the longitudinal resistance $R_{xx}(H)$ and the Hall resistance $R_{xy}(H)$ isotherms measured in HOPG samples show crossings at certain values of applied magnetic field $H \gg H_c$ (Figs. 22 – 24). No such crossings take place for Kish graphite (Figs. 26 and 27).

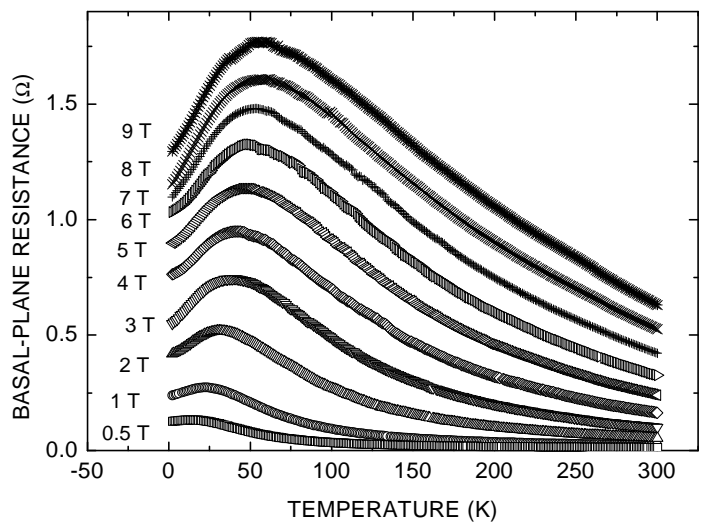

Fig. 20. Basal-plane resistance measured in KISH-1 single-crystalline graphite sample in the high-field limit.

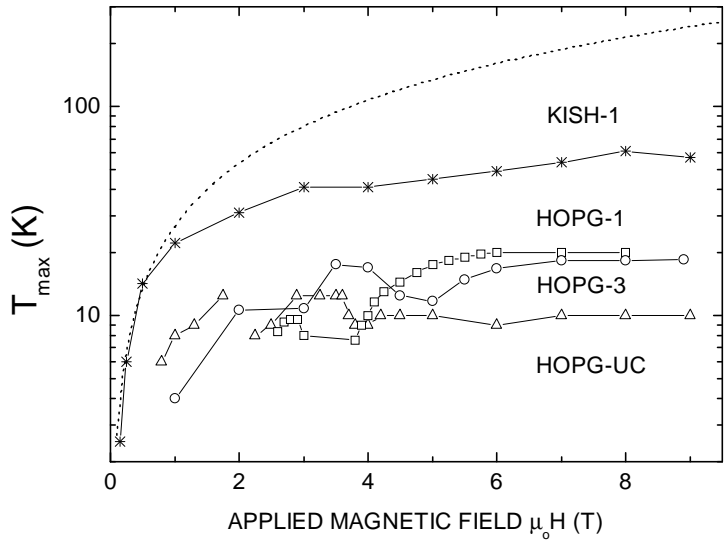

Fig. 21. $T_{max}$ vs. H obtained for several graphite samples; dotted line corresponds to Eq. (6) with $m^* = 0.05 m_0$.

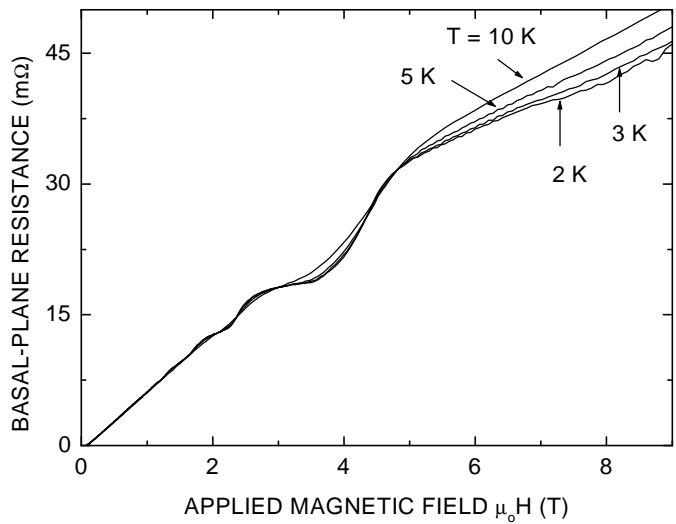

Fig. 22. Basal-plane resistance measured in HOPG-3 sample at four temperatures, demonstrating crossings of the $R_{xx}(H)$ isotherms, i. e. a sequence of the field-driven metal-insulator-metal transitions (or crossovers).

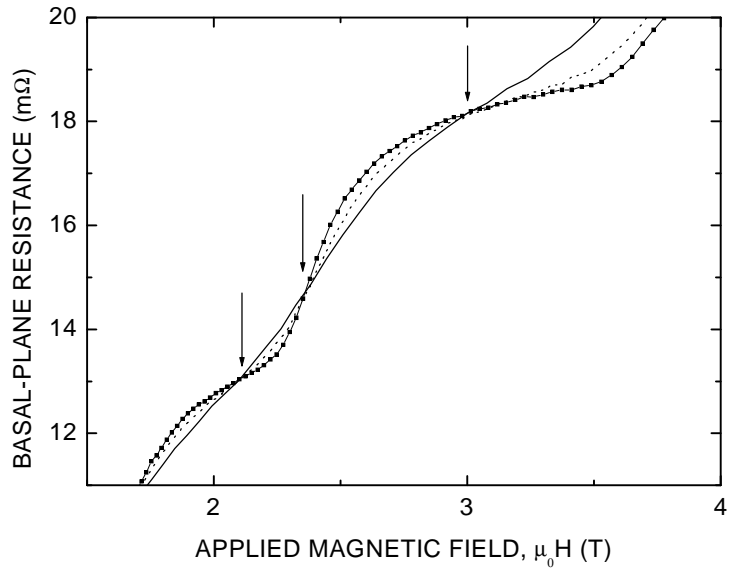

Fig. 23. Detailed view of the $R_{xx}(H)$ isotherms (Fig. 22) crossing; T = 2 K (●), 5 K (dotted line), 10 K (solid line). Arrows indicate magnetic fields at which metal ($dR_{xx}/dT > 0$) – insulator ($dR_{xx}/dT < 0$) or insulator-metal transitions take place.

The crossings in both $R_{xx}(H)$ and $R_{xy}(H)$ isotherms in the Landau level quantization regime have been reported for various 2D systems and attributed to insulator (I) – quantum Hall conductor transitions [81-91]. The scaling analysis, similar to that described above, revealed a non-universal exponent $\kappa = 1/z\nu$ ranging from ~ 0.14 to ~ 0.81 (see Refs. [81-91] and references therein). While the Hall resistance in graphite does not demonstrate well defined plateaus, i. e. there is no unambiguous evidence for the QHE, see Fig. 24, we found a reasonable scaling of the resistance curves in the vicinity of "critical" fields $H_x$ separating metallic and insulating states. Figure 25 exemplifies the scaling found at $\mu_0 H_x = 2.35$ T (Fig. 23). The same scaling exponent $\kappa = 0.1$ was found for all "crossing" fields. We note that QHE can occur also in bulk systems, see e. g. Refs. [92-96].

If the observed metal-insulator-metal transitions at high fields are related to the quasi-2D nature of HOPG, the absence of crossings in $R_{xx}(H, T)$ and $R_{xy}(H, T)$ isotherms measured for the Kish graphite (Fig. 26 and Fig. 27) suggests a rather 3D-like behavior of the single crystal. Such conclusion is supported by the lower anisotropy of the Kish graphite (see above) and c-axis resistance measurements, see Section V.

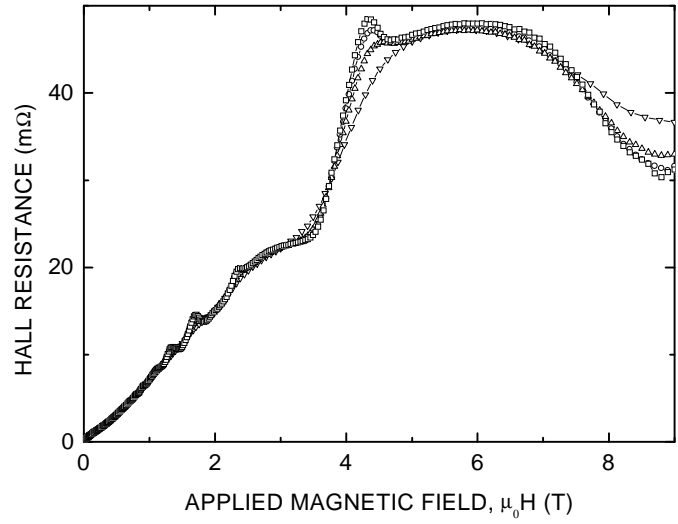

Fig. 24. Hall resistance $R_{xy}(H)$ measured in HOPG-3 sample at T = 2 K (□), 3 K (o), 5 K (Δ), and 10 K (∇).

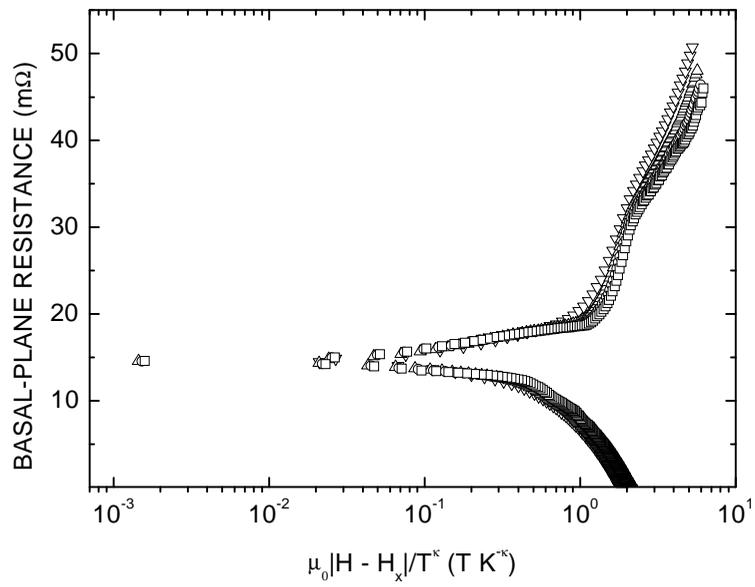

Fig. 25. Basal-plane resistance measured for HOPG-3 at at T = 2 K (□), 3 K (o), 5 K (Δ), and 10 K (∇) plotted vs. the scaling variable, where $\mu_o H_x = 2.35$ T and $\kappa = 0.1$.

On the other hand, the occurrence of the re-entrant metallic state increasing field in both HOPG and Kish graphite samples may be caused by an unique mechanism. It has been proposed in Ref. [5] that the metallic state in the quantum limit is related to the Cooper pair formation. The appearance or re-appearance of superconducting correlations in the regime of Landau level quantization has been predicted [97–103]; for a review see Ref. [104]. The field-induced superconductivity has also been discussed in the context of QHE [66, 67, 105-107]. More recently, the re-entrant high-field superconductivity has been found in a periodic lattice model, considering Hofstadter bands instead of Landau levels [108, 109]. According to the theory of Landau-level-quantization-induced superconductivity [97-104] the increase of the superconducting transition temperature $T_c(H)$ with field results from the increase in a 1D density of states $N_1(0)$ at the Fermi level. In the quantum limit ($H > H_{QL}$) the superconducting critical temperature $T_c(H)$ is given by the equation [104]

$$T_c(H) = 1.14\Omega \exp[-2\pi l^2/N_1(0)V], \qquad (7)$$

where $2\pi l^2/N_1(0) \sim 1/H^2$, $l = (\hbar c/eH)^{1/2}$, V is the BCS attractive interaction, and $\Omega$ is the energy cutoff on V.

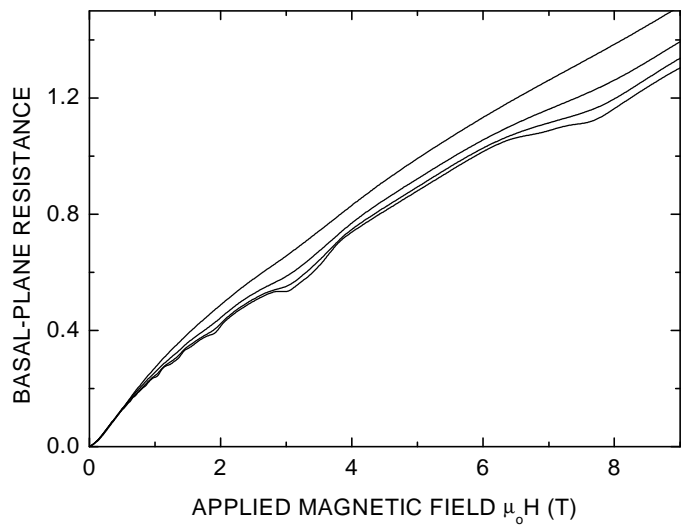

Fig. 26. Basal-plane resistance measured for KISH-1 sample at temperatures (from top to the bottom) T = 20, 10, 5, and 2 K.

The rapid increase of $T_{max}$ with field (see Fig. 21) is in qualitative agreement with Eq. (7). At $H >> H_{QL}$ (in graphite $H_{QL} \sim 4$ T for the hole majority), a saturation in $T_c(H)$ following the reduction of $T_c$ with H is expected; this is also consistent with the observed saturation in $T_{max}(H)$ at high enough fields (Fig. 21).

It is important to note that according to theory a relatively small effective g-factor in graphite $g^* = (m^*/m_0)g \sim 0.1$ ($g \approx 2$) ensures a substantial field interval above $H_{QL}$ where both spin-up and spin-down states are occupied, required for the spin-singlet superconductivity. Also, theory predicts an oscillatory behavior of $T_c(H)$ at $H < H_{QL}$, i. e. with increasing number of occupied Landau levels. In the regime of pronounced Landau level oscillations and for HOPG samples, we observe the non-monotonous $T_{max}$ vs. H behavior, indeed.

We believe that the absence of pronounced $T_{max}$ vs. H oscillations in Kish graphite is caused by its more 3D nature. Indeed, in the 2D case the density of states $N(E_F)$ is a set of delta functions (broadened however by quenched and thermal disorder) corresponding to different Landau levels, and therefore $T_{max}$ oscillates stronger with field in quasi-2D HOPG. At the same time, electron-electron repulsion is more effective in 2D and can act against superconductivity, reducing $T_c$. A $T_{max}(9T) = 62$ K obtained for Kish graphite (see Fig. 21) is much higher than $T_{max}(9\ T) = 11$ K measured in the strongly anisotropic HOPG-UC sample.

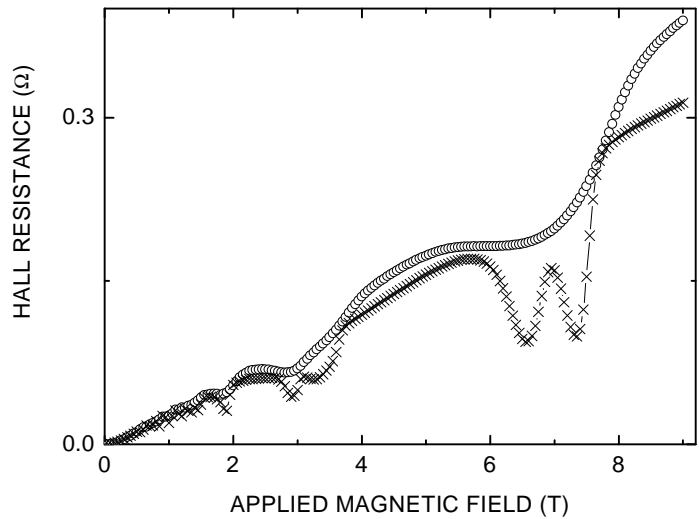

Fig. 27. Hall resistance $R_{xy}(H)$ measured for KISH-1 sample at T = 1.5 K (x) and T = 5 K (o).

Theory [104] predicts that in the superconducting state of 3D samples, the resistance along the applied field direction vanishes and the resistance perpendicular to the field shows a drop. In layered quasi-2D materials the resistance across the conducting planes ($R_c$) can be finite due to weakly- or de-coupled planes. Our measurements of both $R_c(T,H)$ and $R_b(T,H)$ performed on various graphite samples demonstrate that $R_c(T,H)$ is governed by the in-plane resistance [110]. As Fig. 28 demonstrates for the HOPG-UC sample, the c-axis resistance increasing slows down in the vicinity of $T_{max}(9\ T) \approx 11$ K, indeed.

Finally, we note that within the magnetic-catalysis-driven MIT model, the "inverse" MIT, i. e. a metallic resistance behavior below the transition temperature $T_c(H)$ given by Eq. (2) has been predicted for a large enough carrier density [36]. To decide which explanation for the metallic state(s) in the Landau level quantization regime is correct, further work is certainly needed. Interestingly, while the anomalous resistance behavior of graphite observed at $\mu_o H > 20$ T [73-75, 111-117] and attributed to the charge-density-wave (CDW) formation has been a subject of the intense research, no such an attention was given to the resistance behavior at lower fields. In Fig. 29 we construct a "global" phase diagram for Kish graphite, including both our data and recent high-field results reported in Ref. [112]. As Fig. 29 illustrates, there are alternating "metallic" and "insulating" H-T domains whose origin remains to be clarified.

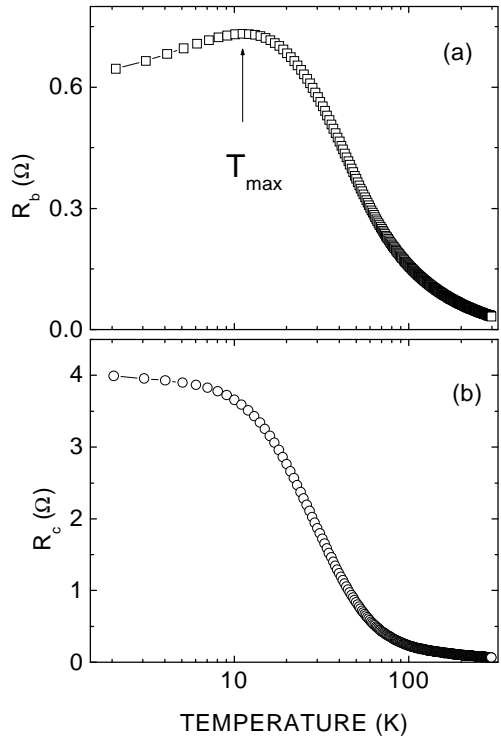

Fig. 28. Basal-plane (a) and c-axis (b) resistance measured for HOPG-UC sample in magnetic field $\mu_o H = 9$ T applied parallel to the sample c-axis.

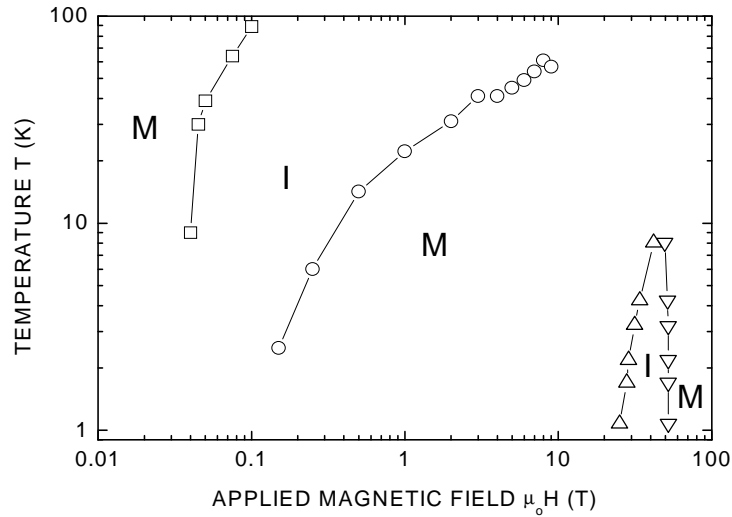

Fig. 29. H-T phase diagram constructed for Kish graphite, where M and I stand for "metallic" and "insulating" states, respectively; (□) $T_{min}(H)$ obtained from the data of Fig. 6 and (o) $T_{max}(H)$ given in Fig. 21; (Δ) and (∇) data points, taken from Ref. [112] attributed to the metal-CDW and CDW-metal transitions, respectively.

## V. THE INTERLAYER MAGNETORESISTANCE OF GRAPHITE: COHERENT AND INCOHERENT TRANSPORT

In a recent work we have showed that the metallic-like behavior of the interlayer c-axis resistivity $\rho_c(T,B)$ of HOPG and Kish graphite samples is directly correlated to that of the longitudinal resistivity of the basal planes $\rho_b(T,B)$ [110]. These results also indicate that the intrinsic temperature dependence of $\rho_c(T,0)$ is very likely semiconducting one and should be observed for a sample with a low density of defects. We note that lattice defects may not only affect the transport as scattering centres but they may contribute to enhance the electronic coupling between layers giving rise to a quasi-3D electronic spectrum with coherent transport along the c-axis. The situation may be even more tricky since defects can produce locally superconducting-like domains and therefore produce a strong decrease of the resistivity with temperature. In this case, more defects do not necessarily mean larger resistivity.

These results [110] cast also doubts about the estimate of ~ 0.3 eV for the interlayer transfer integral used in the literature [32]. This is not surprising since neither the electron-electron interaction nor charge fluctuations [118] were taken into account. In principle, for such a large interlayer transfer we would expect coherent transport for the interlayer magnetoresistance at low temperatures, even for perfect, defect-free samples. One possible way to test coherent transport across the graphite layers is given by the measurement of a peak in the angle dependence of the c-axis magnetoresistivity at fields parallel to the layers [119].

This peak should be absent for incoherent interlayer transport and should be seen if the inequality $\omega\tau > 1$ holds, where $\omega$ is the cyclotron frequency and $\tau$ the relaxation time of the carriers. Coherent transport means therefore that band states, extending over many layers, and a 3D Fermi surface can be defined. In the other case, incoherent transport is diffusive and neither a 3D Fermi surface can be defined nor the Bloch-Boltzmann transport theory is applicable [119]. Taking the small Fermi energy ~ 200 K of graphite we would expect to observe a peak as a function of angle for magnetic fields less than ~10 T in case the interlayer transport is coherent.

High-resolution field-angle dependent experiments around the direction parallel to the planes have been recently done [120]. The measurements show a weak coherent peak in the $\rho_c$ resistivity for Kish graphite but not for good quality HOPG samples. Figure 30 shows the angle dependence of $\rho_c$ at a temperature of 2 K at different applied fields. The Kish graphite sample has a relatively broad width of the rocking curve (FWHM ~1.6°) and a very low electrical resistivity.

This sample shows clearly a broad peak in $\rho_c$ at 90° field direction, i. e. at a field applied parallel to the planes, see Fig. 30 (b). In contrast, two HOPG samples with FWHM = 0.4° (close box) and 0.6°(+), see Fig. 30 (b), show no maximum. The asymmetry seen in the angle dependence in some of the samples is in part due to the small misalignment of the plane of the sample to the rotation axis and to the lack of crystal perfection. Our results indicate that, as expected from theoretical considerations, coherent transport is observed in a sample which is apparently more disordered and therefore has better coupling between planes. This sample (KISH-2) shows a 3D-like behavior and also much lower resistivity at low temperatures in comparison with the HOPG samples.

## VI. MAGNETO-THERMAL CONDUCTIVITY IN HOPG NEAR THE METAL-INSULATOR TRANSITION

The thermal conductivity $\kappa(T,H)$ of highly oriented graphite samples and its dependence with magnetic field has been measured at least twice long time ago [121,122]. Actually and due to the layered structure of the material one would expect deviations from simple Fermi liquid theoretical predictions. In fact, some experimental results remained unexplained without attracting great interest in the scientific community. Also, the sensitivity of those measurements was probably not good enough to observe any anomalous behavior at low fields. It has been shown that at temperatures below ~ 4.2 K, $\kappa$ decreases with field up to a field 0.1…0.2 T; above this field it saturates to an approximately constant value [121,122]. This saturation has been interpreted as due to the freezing out of the thermal conduction by electrons with field [122]. However, the relatively large amplitude of the quantum oscillations (e.g. at fields H ~ 3.5…4T) observed at similar temperatures and within the saturation level [121] does not appear to be compatible with the simple interpretation that the conduction electrons are frozen out and do not contribute to the thermal transport.

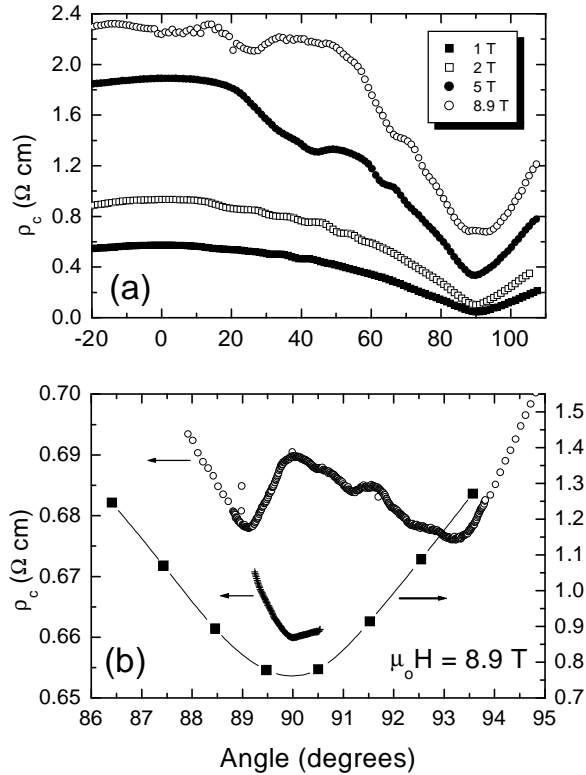

Fig. 30. (a) Angle dependence of the c-axis resistivity $\rho_c$ at T = 2K for KISH-2 crystal at different applied fields; (b) open circles show the same data as in (a) for $\mu_oH = 8.9$ T but taken with higher angle resolution around the field direction parallel to the planes. Note the broad maximum with weak oscillations around 90°. For comparison we show in the same figure the results for two HOPG samples ((+): HOPG-3, (◊): HOPG from AC) at the same temperature.

On the other hand, the quantum oscillations as a function of field observed in $\kappa(B)$ are in phase with those found in the electrical resistivity. The electrical resistivity increases monotonically with field (see for example Fig. 22) in such a way that and following the Wiedemann-Franz (WF) law, we would expect a vanishing contribution of the electrons at high fields and a small oscillation amplitude due to the Landau-level crossing the Fermi energy. Therefore, the origin of the "saturation" of the thermal conductivity at $\mu_0H > 0.2$ T at low temperatures as well as the validity of the WF-law in graphite do not seem to be really clarified within the Fermi-liquid theory.

The behavior of the thermal conductivity with field can be different from that obtained for the electrical conductivity because elastic and inelastic processes with low-wave vector at the Fermi surface play a more significant role in $\kappa$ than in $\rho$. Taking also into account the magnetic-field driven metal-insulator transition found in graphite, the measurement of the field dependence of $\kappa$ at low fields and with higher resolution than it was performed before appears to be necessary. The measurements recently done [123] indicate a non-monotonous behavior of $\kappa$ with field, at fields of the order of the "critical" field $\mu_0 H_c \sim 0.1$ T found in the electrical resistivity measurements. This result appears to violate the WF law.

Like in conventional metals, the total thermal conductivity in graphite is given by the sum of the contributions due to phonons, $\kappa^{ph}$, and the conduction electrons, $\kappa^{el}$. Therefore assumptions are necessary to separate the relatively large phonon contribution from the total thermal transport in graphite. Unfortunately, there is no simple method that allows a straightforward separation of $\kappa^{el}$ at all temperatures in graphite. One method is based on a phenomenological description of the field dependence of the longitudinal thermal conductivity $\kappa(T, H)$, which account for the scattering of quasiparticles by vortices in superconductors.

Considering that the thermal conductions due to phonons does not depend on magnetic field, one may assume a field dependence of the thermal conductivity for graphite of the form $\kappa(T, H) = \kappa^{ph}(T) + [\kappa^{el}(T)/(1+ \text{ß}(T) H^n)]$ where, in general, ß(T) is proportional to the zero-field electronic mean-free-path and the exponent n can be related to the nature of the electronic scattering. It is clear that this simple separation can be used at temperatures where the quantum oscillations in $\kappa$ are negligible. In our measurements we found that at T > 20 K and for fields parallel to the c-axis, $\kappa$ decreases with field and because at those temperatures the quantum oscillations are rather small we obtain a good fit to the data with the above simple approximation. Using this method, however, we obtained neither a quantitative nor a qualitative agreement with the predictions for $\kappa^{el}(T)$ from the WF law taking into account the measured electrical conductivity. For example, at T = 34 K we have measured $\kappa \sim$ 690W/Km, and from the fit of its field dependence we obtained $\kappa^{ph} \sim 678$W/Km, $\kappa^{el} \sim$ 11.7W/Km, ß $\sim 0.175 T^{-1}$ and n = 1.03 ± 0.29. At T = 24K we obtain $\kappa \sim 470$W/Km and from the fit $\kappa^{ph} \sim 458$W/Km, $\kappa^{el} \sim 12.2$W/Km, ß $\sim 0.39 T^{-1}$ and n = 0.85 ± 0.10. The absolute value of the measured thermal conductivity agrees within a factor two to that measured by Ayache [121]. Also our estimates for the ratio $\kappa^{el}/\kappa^{ph}$ for those temperatures are in good agreement with those from Ref. [121]. From the electrical resistivity measurements on the same sample we obtain $\rho_b(34K) \sim 0.57$ µΩm and $\rho_b(24K) \sim 0.51$ µΩm. With the WF law we obtain then $\kappa^{el} \sim 1.5$ and 1.1W/Km at 34 K and 24 K, respectively, in clear disagreement with the values estimated above. This result is in contrast to Ayache's work. One would tend to ascribe this disagreement to the absolute value obtained for the electrical resistivity which may be influenced by geometrical effects. However, within a factor of two, similar values of resistivity were obtained for other pieces of the same sample and for other HOPG samples of similar quality. We estimate the error in the absolute value of the resistivity to be less than a factor two.

A more systematic work is necessary to check whether a quantitative agreement with the WF law is achieved for more quasi-3D samples due to their increasing defect density and consequently lower resistivity.

In what follows we would like to compare the field dependence of the thermal conductivity with that of the electrical resistivity at low temperatures. Figure 31 shows the reduced thermal conductivity as a function of magnetic field applied perpendicular to the basal planes and at T = 6 K. The thermal conductivity decreases ~ 5% with a field of 0.05 T. At this field κ shows a drastic change of slope with a clear minimum and it increases with field up to ~ 0.25 T where a maximum is observed. We note first that the minimum is at a similar field as the critical field obtained from the scaling of the electrical resistivity (see Fig. 13). The field dependence κ(H) is in clear contrast to the monotonous dependence observed in the electrical resistivity. In Fig. 31 we show the magnetic field dependence of the in-plane resistivity $\rho_{xx}$ at 2K and 10 K for the same HOPG-AC sample and at 5 K for a HOPG-3 sample. Clearly, there is a violation of the WF law with magnetic field. The anomalous behavior of κ(H), specially the increase with field, at low fields, starts to be clearly discernible at T ~10K [123] in perpendicular fields.

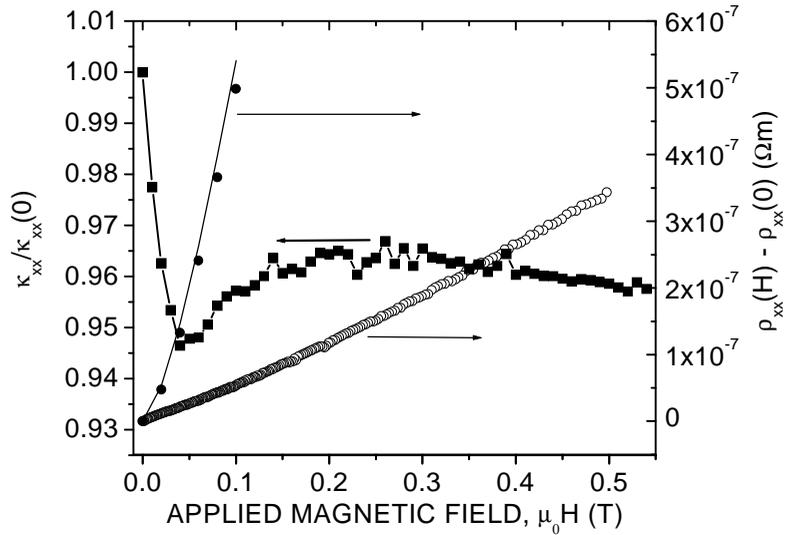

Fig. 31: Field dependence of the normalized thermal conductivity at T = 6 K of a HOPG-AC sample. In the same figure we show the absolute change of the longitudinal electrical resistivity for the same sample at 2K (continuous line) and at 10K (close circles), and for a HOPG-3 sample at 5K (open circles). Note the clear minimum at ~0.05T in κ(H).

It is tempting to relate the observed behavior in the thermal conductivity to the magnetic-field driven metal-insulator transition. We note first that the anomalies observed in the thermal conductivity occur at lower temperatures than the $T_{min}(H)$ obtained for similar samples (see Fig. 6). According to a recent work [124] the arising of a gap due to the magnetic catalysis phenomenon, recently proposed as a possible origin for the observed metal-insulator transition in graphite [35], would have a drastic effect in the magnetic field dependence of the thermal conductivity. In particular, a kink in $\kappa(B)$ is predicted at a temperature dependent field $H_k(T)$. In the case of weak coupling between the quasiparticles, a magnetic field $H_k(T)$ would trigger an increase in the fermion mass (through a coupling between electrons and holes) generating a kink in $\kappa(H)$. The thermal conductivity would first decrease with field at $H < H_k(T)$ and would remain constant at larger fields or even increase near the transition at $H > H_k(T)$. The behavior shown in Fig. 31 resembles qualitatively this prediction. However, we stress that the critical field $H_k(T)$ obtained from the thermal conductivity is not the one obtained from the minimum in the electrical resistivity. It is unclear at present the meaning of this critical field and its possible relationship with the MIT.

## VII. EVIDENCE FOR SUPERCONDUCTING AND FERROMAGNETIC CORRELATIONS: MAGNETIZATION STUDIES

### VII.1. Superconducting and ferromagnetic instabilities in graphite

The possible existence of superconducting correlations in graphite, as magnetoresistance measurements suggest, led us to a detailed study of magnetic properties of graphite [6, 125]. In particular, it has been found that a characteristic feature most of "virgin" HOPG samples is the occurrence of magnetization hysteresis loops M(H) of superconducting (SC) - and ferromagnetic (FM) - type for magnetic field applied parallel (H ∥ c) and perpendicular (H ⊥ c) to the sample c-axis, respectively.

Figure 32 shows SC-like hysteresis loops for HOPG-1, HOPG-2, HOPG-3 samples and FM-like loop for HOPG-UC, all obtained with H ∥ c-axis. In this geometry, the major signal is due to orbital diamagnetic contribution [31] $M = -\chi_\parallel H$ which has been subtracted.

However, M(H) hysteresis loops measured in H ⊥ c geometry, see Fig. 33, are typical for ferromagnets. It should be emphasized that M(H) presented in Fig. 33 were obtained without subtraction of any background signal. Our studies revealed also that a diamagnetic signal measured at high enough fields and in the H ⊥ c geometry results from the sample misalignment.

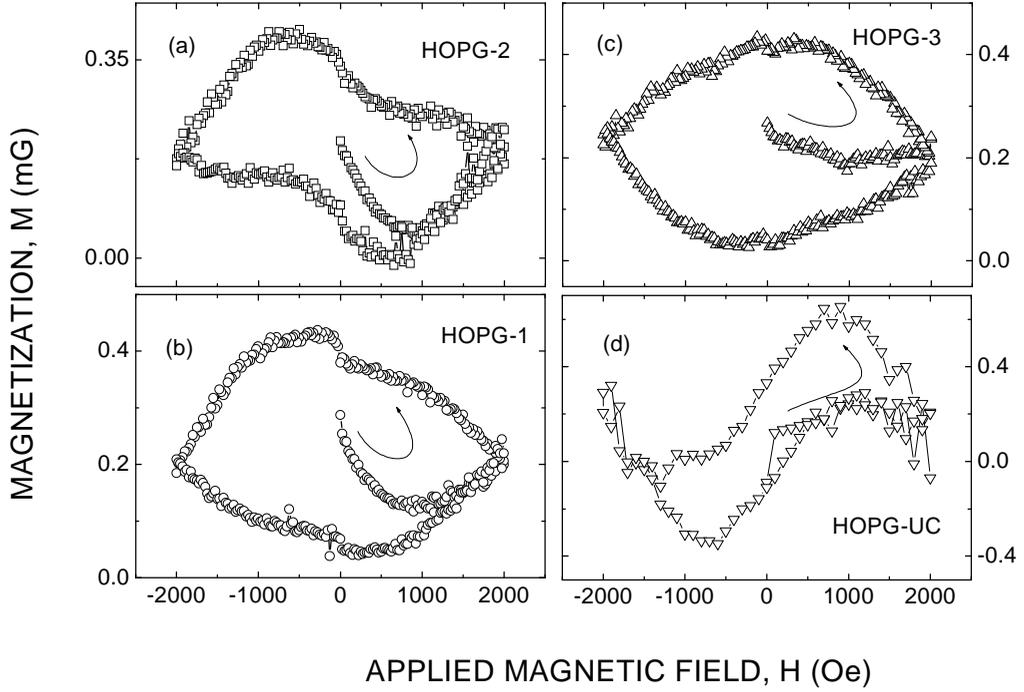

Fig. 32. Magnetization hysteresis loops obtained in magnetic field H ∥ c-axis at T = 300 K after subtraction of the diamagnetic background signal $M = -\chi H$, with $\chi = 4.21 \cdot 10^{-2}$ mG/Oe (HOPG-1), $\chi = 4.185 \cdot 10^{-2}$ mG/Oe (HOPG-2 [6]), $\chi = 4.314 \cdot 10^{-2}$ mG/Oe (HOPG-3), and $\chi = 4.185 \cdot 10^{-2}$ mG/Oe (HOPG-UC).

As can be seen in Fig. 33, M(H) saturates at H ≥ 1 kOe, and the saturation magnetization $M_s$ in HOPG-UC sample is about 4 times larger than that obtained for other three HOPG samples. Comparing the results presented in Figs. 32 and 33, one arrives to the conclusion that the stronger ferromagnetism in the HOPG-UC sample masks or suppresses the superconducting behavior seen for HOPG-1, HOPG-2, and HOPG-3 samples. Actually, our assumption that MIT transition (see Fig. 3) might be a Bose metal–insulator transition would imply that the ferromagnetism coexists with superconducting correlations. An unambiguous evidence for the coexistence of ferromagnetism and superconductivity has recently been reported for graphite-sulfur composites [9].

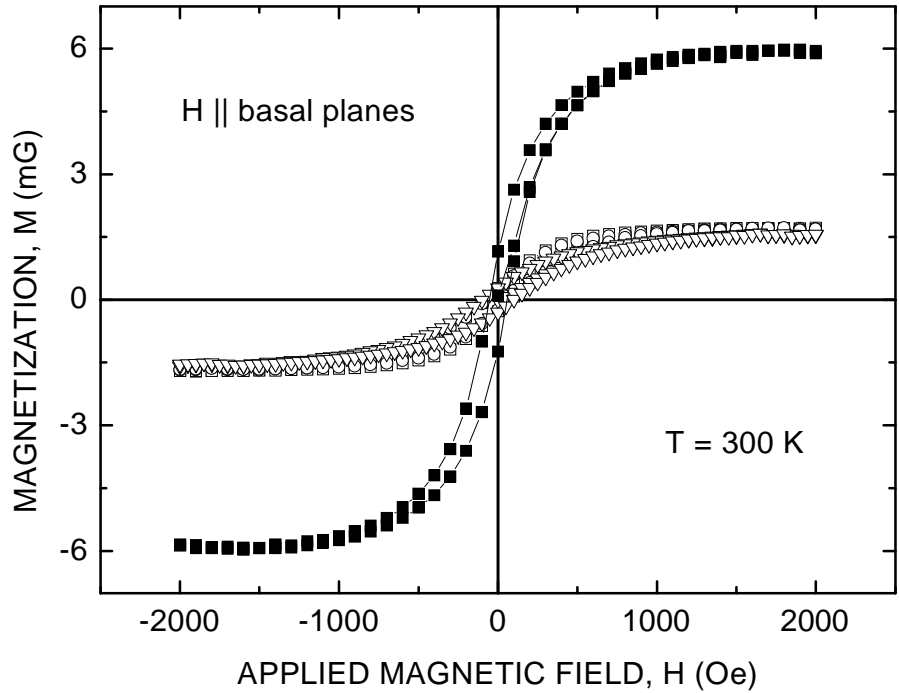

Fig. 33. Ferromagnetic hysteresis loops measured at T = 300 K for HOPG-1 (o), HOPG-2 (■), HOPG-3 (∇), and HOPG-UC (□) with magnetic field applied parallel to the sample basal planes.

It appears that both width of ferromagnetic loops and $M_s$ increase under sample annealing in low vacuum, see Fig. 34. Figure 35 (a,b) demonstrates that the annealing-induced increase of the ferromagnetism masks the superconducting behavior seen in the virgin sample for H || c. We stress, that the sample transport properties are not affected by the annealing indicating that the magnetization change is a local effect.

An unambiguous evidence for the local superconductivity occurrence at high temperatures has been reported for graphite-sulfur (CS) composites [7-9], suggesting that SC-like hysteresis loops measured in various graphite samples at room temperature are related to local superconductivity. We note that in both pure and CS composites the resistance remains finite down to the lowest measured temperature indicating that superconducting "grains" are isolated from each other or coupled into finite clusters.

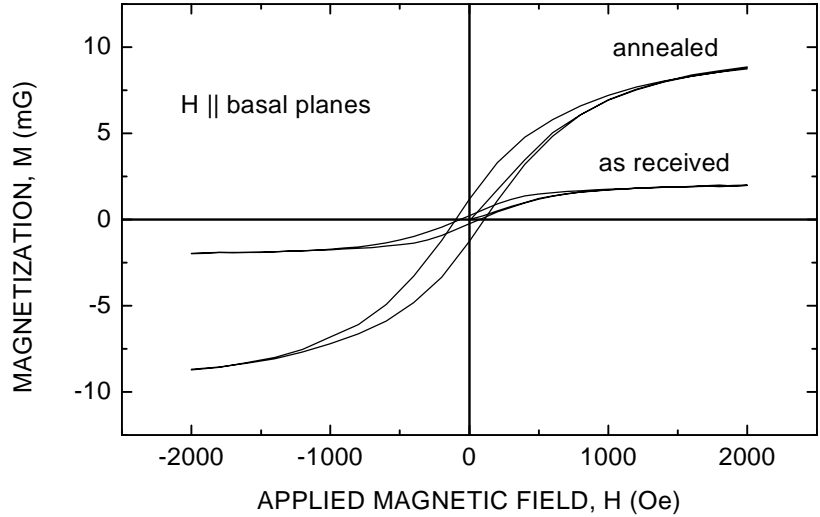

Fig. 34. M(H) measured for HOPG-2 sample before ("as received") and after annealing at T = 800 K for 2 hours.

VII.2. Local superconductivity in graphite-sulfur composites

Figure 36 presents temperature dependencies of the magnetization $M(T,H) = m(T,H)/V$ (m is the sample magnetic moment and V is the sample volume) measured in as-received sample (labeled here as A) at applied fields H = 10 Oe and H = 100 Oe. The magnetization data corresponding to the zero-field-cooled (ZFC) regime, $M_{ZFC}(T)$, were taken on heating after the sample cooling at zero applied field, and the magnetization in the field-cooled on cooling (FCC) regime, $M_{FCC}(T)$, was measured as a function of decreasing temperature in the applied field. Figure 36 demonstrates a pronounced difference between $M_{ZFC}(T)$ and $M_{FCC}(T)$ which occurs with the temperature decreasing. Figure 37 gives a detailed view of the data obtained for H = 100 Oe in a vicinity of the $T_c(H = 100\ Oe) = 33$ K below which a departure of $M_{ZFC}(T)$ from $M_{FCC}(T)$ takes place. As can be seen from this plot, both $M_{ZFC}(T)$ and $M_{FCC}(T)$ become more diamagnetic at $T < T_c(H)$.

Such magnetization behavior is characteristic of superconductors: the enhancement of the diamagnetism below the superconducting transition temperature $T_c(H)$ originates from the screening supercurrents and the Meissner-Ochsenfeld effect of magnetic flux expulsion (ZFC regime) whereas in the FCC regime the flux is trapped and the magnetization remains nearly temperature independent due to vortex pinning. It can also be seen in Fig. 36 that as the applied field increases, the normal state orbital diamagnetism of graphite overcomes a positive contribution to the magnetization, which can be due to both intrinsic weak ferromagnetism of graphite and magnetic impurities, resulting in a negative total magnetization above $T_c$.

Figure 38 presents magnetization M(H) measured at T = 6 K after cooling the sample from 300 K to the target temperature in a zero applied field; the occurrence of magnetic hysteresis characteristic of type-II superconductors with vortex pinning can be seen in the CS even without subtraction of the orbital diamagnetic contribution.

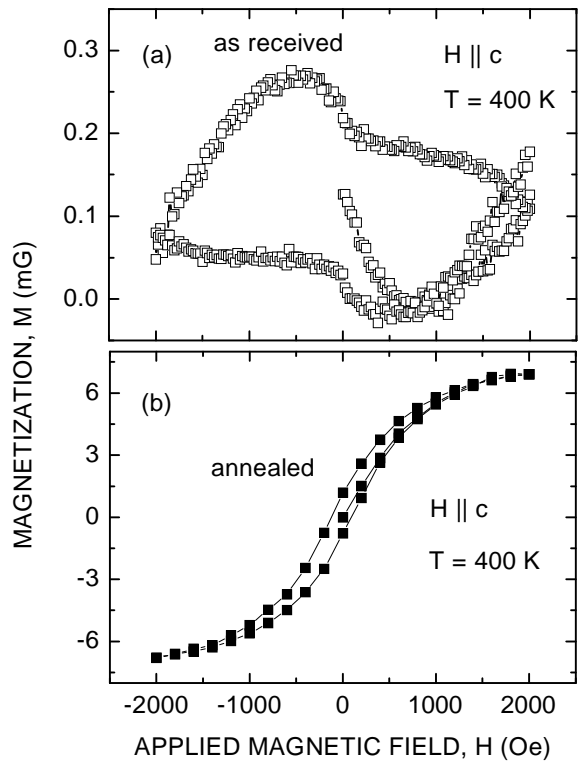

Fig. 35. The annealing-induced transformation of superconducting-like magnetization hysteresis loop (a) to the ferromagnetic-like (b) obtained after annealing the as-received sample for 2 hours at T = 800 K inside of the SQUID magnetometer's oven. The diamagnetic background signals $M = -\chi H$ were subtracted: $\chi = 3.147 \cdot 10^{-2}$ mG/Oe (a) and $\chi = 3.1 \cdot 10^{-2}$ mG/Oe (b).

It appears, that a loss of sulfur leads to a considerable decrease in the superconducting response but leaving $T_c(H)$ nearly unchanged [7]. The magnetization data obtained for the same sample after it lost ~ 4 wt % of sulfur (labeled here as CS-1) are given in Figs. 39 - 41.

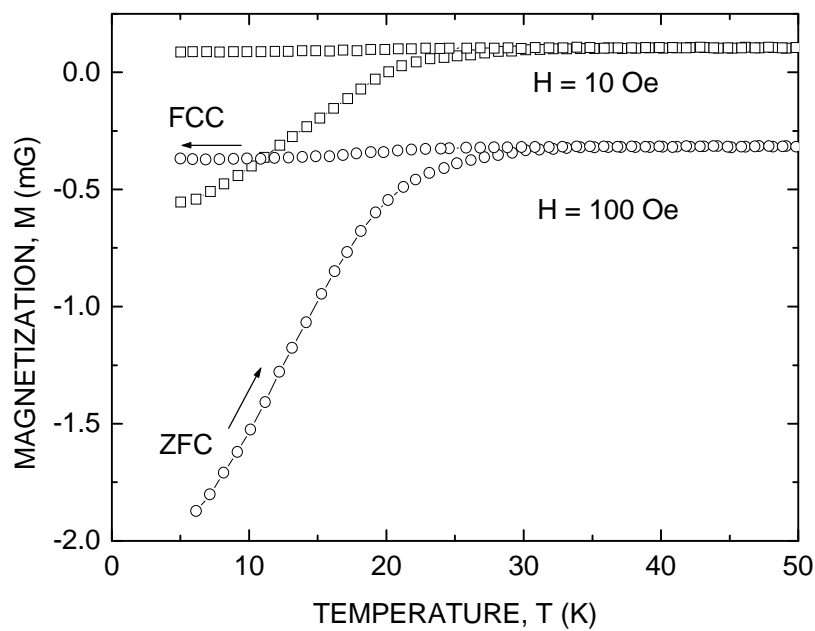

Fig. 36. Temperature dependencies of magnetization M(T) measured in CS composite in zero-field-cooled (ZFC) and field-cooled on cooling (FCC) regimes at two applied fields; 10 Oe and 100 Oe.

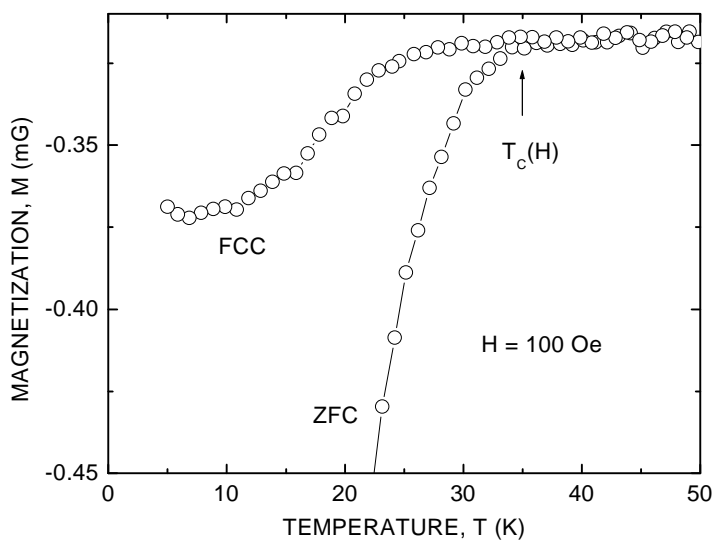

Fig. 37. Enlarged view of the superconducting transition (see Fig. 34) recorded at H = 100 Oe.

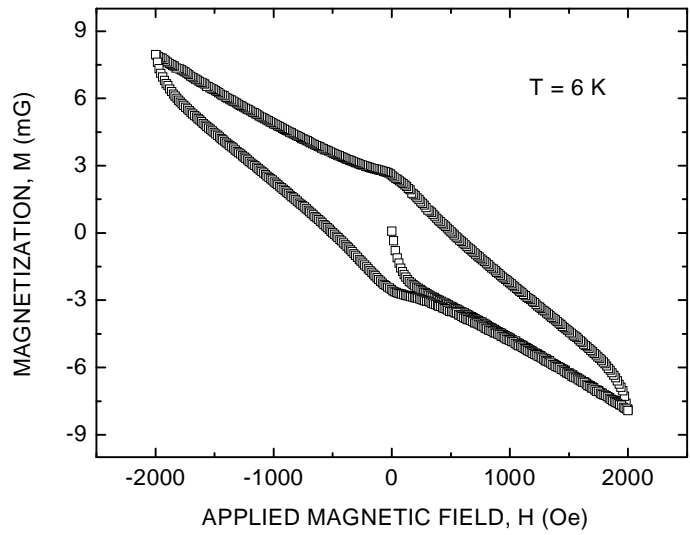

Fig. 38. Magnetization hysteresis loop M(H) measured in CS composite at T = 6 K.

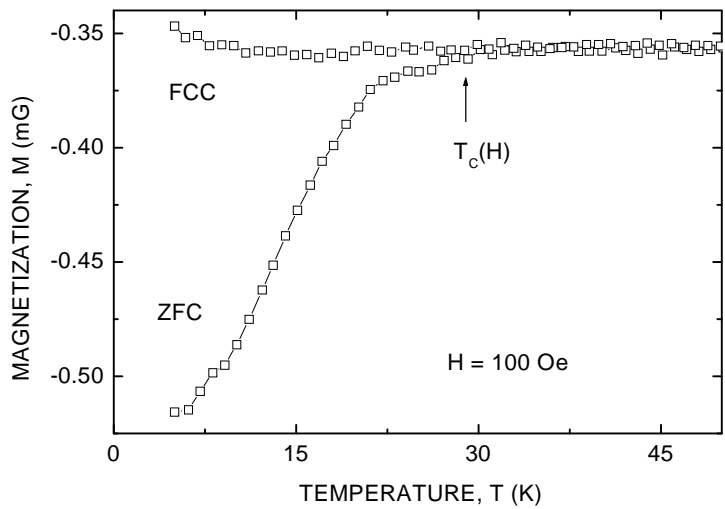

Fig. 39. M(T) measured for the sample CS-1.

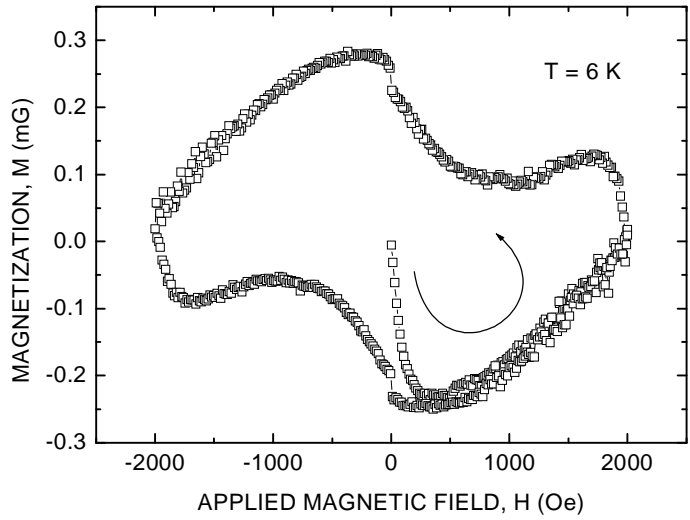

Fig. 40. Superconducting magnetization hysteresis loop M(H) obtained after subtraction of the diamagnetic background signal $M = -\chi H$ with $\chi = 3.5 \cdot 10^{-3}$ mG/Oe in the CS-1 sample at 6K.

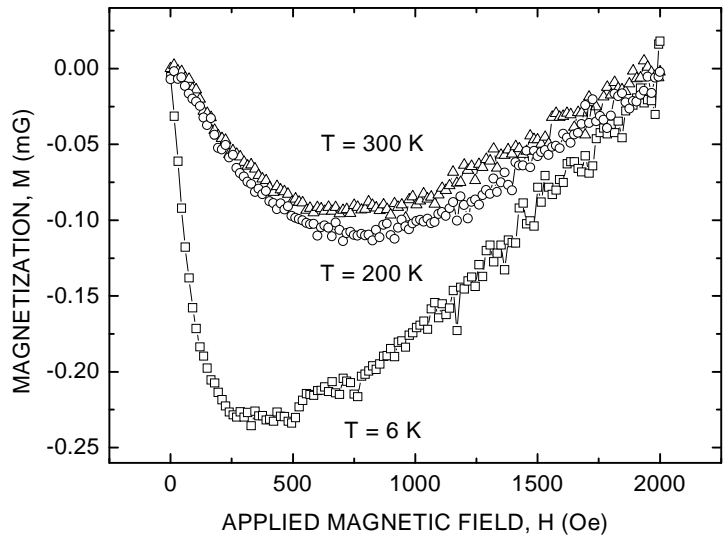

Fig. 41. Virgin magnetization obtained for the CS-1 sample below and above $T_c(0) \sim 30$ K after subtraction of the diamagnetic background signal $M = -\chi H$ with $\chi = 3.5 \cdot 10^{-3}$ mG/Oe (T = 6 K), $\chi = 2.91 \cdot 10^{-3}$ mG/Oe (T = 200 K), and $\chi = 2.6 \cdot 10^{-3}$ mG/Oe (T = 300 K).

Figure 41 shows virgin magnetization curves M(H) obtained for the CS-1 sample below and well above the superconducting transition temperature $T_c(H = 0) \sim 30$ K. As can be seen from that plot, the Meissner-like portion of the M(H), viz. where magnetization decreases with field, persists up to room temperature, at least, suggesting the existence of superconducting "grains" at $T \gg T_c$. Then, $T_c(H)$ measured in CS and CS-1 samples (Fig. 42) can be associated with a formation of clustres of coupled "grains" via Josephson tunneling or the proximity effect. Indeed, we found that $H(T_c)$ for CS sample can be best described [7] by the power law:

$$H = H^*(1-T_c/T_{c0})^{3/2} \qquad (8)$$

in the vicinity of $T_{c0} = 35$ K, and by the equation:

$$H = H_0 \exp(-T_c/T_0) \qquad (9)$$

below a reduced temperature $T/T_{c0} \sim 0.8$, where $\mu_0 H_0 = 5$ T and $T_0 = 7$ K, see Fig. 42. Equations (8) and (9) imply that $T_c(H)$ can be accounted for by the existence of a breakdown field $H_b(T)$ which destroys the superconductivity induced by a proximity effect [126-128]. According to the theory [128], $H_b(T)$ for normal-metal-superconductor structures saturates in the limit $T \to 0$ to the value $H_b(T = 0) \approx 0.37 H_0$. Taking $\mu_0 H_0 = 5$ T, one gets $\mu_0 H_b(T = 0) = 1.85$ T which agrees with the experimentally determined field $\mu_0 H = 1$ T at which the superconducting response vanishes, see Fig. 43. Similar results were obtained in Ref. [9] for the graphite-sulfur system with $T_c(0) = 9$ K.

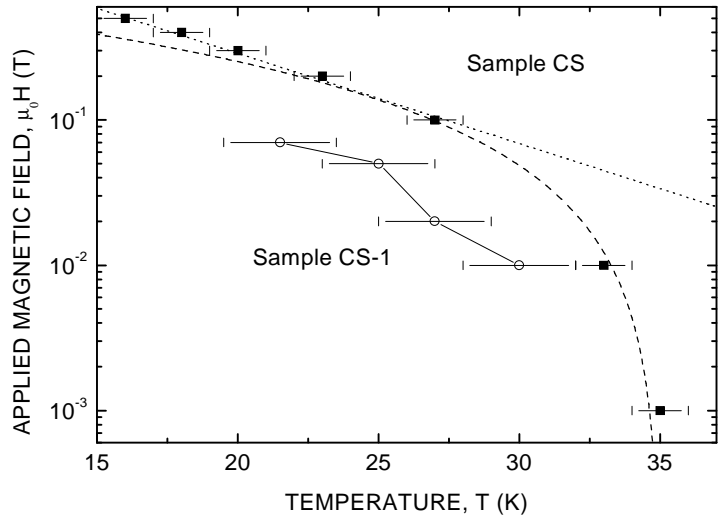

Fig. 42. $H(T_c)$ measured for samples CS and CS-1. Dashed and dotted lines are obtained from Eq. (8) and Eq. (9) with the fitting parameters $T_{c0} = 35$ K, $\mu_o H^* = 0.9$ T, $T_0 = 7$ K, and $\mu_o H_0 = 5$ T.

If our interpretation of the $H(T_c)$ is correct, one can speculate that the main role of sulfur is to couple preexisting superconducting "grains" forming clusters of different size [129]. Then, the reduced superconducting response in the sample CS-1 as compared to that obtained for the sample CS can be understood assuming that the size of SC clusters in sample CS-1 is smaller than that in sample CS.

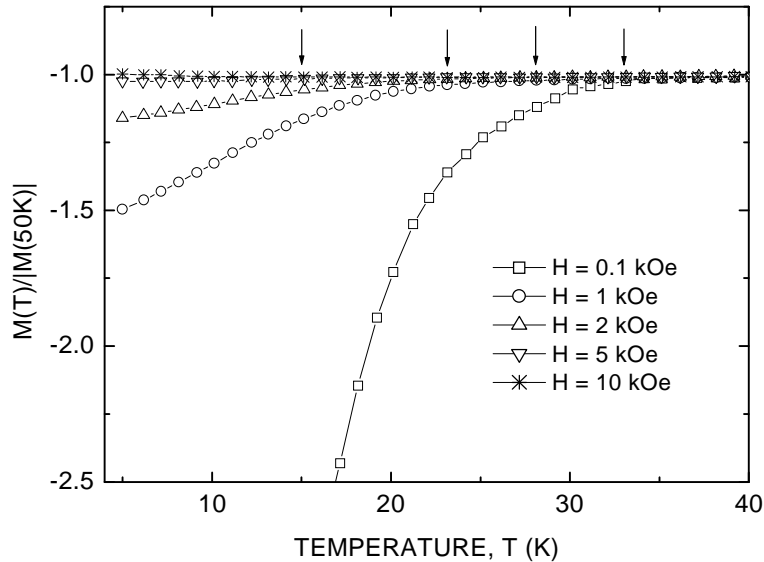

Fig. 43. Normalized ZFC magnetization measured in the sample CS at various applied fields. Arrows denote the superconducting transition temperature $T_c(H)$.

Before closing this chapter we would like to comment briefly on the possible origin of the ferromagnetic behavior of graphite.

VII.3. Possible origin of ferromagnetism in graphite.

*(1) Magnetic impurities.* A trivial explanation for the observed ferromagnetic behavior in graphite would be the presence of magnetic impurities, e. g. Fe. This issue has been addressed and carefully studied in Ref. [125]. Magnetization measurements together with Particle Induced X-ray Emission (PIXE) analysis using a 2 MeV proton beam performed on several HOPG, Kish and natural graphite crystals with a Fe contents ranging from ~ 1 μg/g to 3790 μg/g unambiguously demonstrated that the ferromagnetism in graphite cannot be explained by magnetic impurities.

*(2) Grain boundaries and other topological defects.* It is expected that the electronic structure of graphite may differ considerably on carbons located within a graphite sheet and those at its edge [130, 131]. Moreover, such localized edge states have been shown to depend sensitively on the geometry of the edge [132]: while "zigzag" edges show such states, "armchair" edges do not. Thus, "zigzag" edge states may lead to an increase of the density of states at the Fermi level. This has been experimentally confirmed in STM work on HOPG samples [133]. If such edge states occur at high density, a ferrimagnetic spin polarization may result. A strong enhancement of the Curie-like paramagnetic contribution was predicted at low temperatures, see Ref. [134] and references therein, and this has been experimentally confirmed [135]. Recently, the mechanisms of magnetism in stacked nanographite were theoretically studied [136]. This work obtained an antiferromagnetic solution for A-B-type stacking only.

One may doubt, however, that the edge states could be the origin for the ferromagnetism in graphite samples of a macroscopic size. Indeed, in nanographites these states provide an enhanced paramagnetism and their size ~ 2 nm is smaller than the in-plane correlation length D ~ 10 nm of our graphite samples; the influence of the edge states to the electronic properties of a graphene layer should decrease with D. However, small-size effects and the thermal energy could also preclude the formation of a stable ferromagnetic ordering in nanographite. It is also unclear whether correlation effects between edge states in macroscopic graphite samples with an in-plane correlation length not more than 5 times the size of the reported nanographite samples would not be enough to stabilize a ferromagnetic order.

On the other hand, theoretical studies suggest that the presence of strong topological disorder within graphene planes will stabilize ferromagnetism and frustrate antiferromagnetic order [71].

*(3) Coulomb-interaction-driven ferromagnetic ordering.* The occurrence of a ferromagnetic moment associated with the magnetic-field-induced excitonic gap $\Delta(T, H)$ and the metal-insulator transition accompanied by the time-reversal symmetry breaking has been proposed in Ref. [35]. Supporting theoretical predictions of [35], conduction electron spin resonance measurements [34] performed on HOPG samples using S(4 GHz), X (9.4 GHz), and Q(34.4 GHz) microwave bands demonstrated the occurrence of internal ferromagnetic field and suggested its intimate coupling to the magnetic-field-driven MIT.

As an alternative to the "magnetic catalysis" [35] phenomenon, noting that a gap opening due to spontaneous time-reversal invariance broken (which implies magnetic ordering at H = 0) has been suggested in Ref. [137] taking into account transfer integrals between next-nearest-neighbor sites on the hexagonal lattice of 2D graphite. It has been argued [137] that in this case the electronic gap is related to the integer QHE. Because of the thermally-activated metallic resistance behavior in graphite (see III.4), that possibility has to be explored in more details, probably taking into account dissipation effects [138].

Relatively large values of the Coulomb coupling constant $r_s = 5...10$ (see III.5.) may indicate a proximity of the electron system in graphite to the Wigner crystallization and therefore to the ferromagnetic ground state. It is interesting to note that Hartree-Fock calculations performed by Wigner [139] for Cs ($r_s = 6.5$) yielded ferromagnetic ordering.

More recently, reported in Ref. [34] internal ferromagnetic field in graphite has been attributed to a gapless spin-1 neutral collective mode [64] discussed in the context of RVB state.

Acknowledgements

The experimental work presented in this chapter has been possible with the support of the following institutions and grants: FAPESP, CNPq, CAPES, DFG ES 86/6-1 and the DAAD.

We gratefully acknowledge the interest and discussions with D. V. Khveshchenko, R. Höhne, I. A. Shovkovy, V. A. Miransky, E. V. Gorbar, F. Guinea, V. P. Gusynin, C. Tejedor, F. Flores, E. Ferrer, A. Abrikosov, and K.-H. Müller.

REFERENCES

[1] N. B. Hannay, T. H. Geballe, B. T. Matthias, K. Andres, P. Schmidt, and D. MacNair, Phys. Rev. Lett. 14 (1965) 225.
[2] I. T. Belash, A. D. Bronnikov, O. V. Zharikov, and A. V. Palnichenko, Solid State Commun. 64 (1987) 1445.
[3] A. F. Hebard, M. J. Rosseinsky, R. C. Haddon, D. W. Murphy, S. H. Glarum, T. T. M. Palstra, A. P. Ramirez, and A. R. Kortan, Nature 350 (1991) 600.
[4] K. Tanigaki, T. W. Ebbesen, S. Saito, J. Muzuki, J. S. Tsai, Y. Kubo, and S. Kuroshima, Nature 352 (1991) 222.
[5] Y. Kopelevich, V. V. Lemanov, S. Moehlecke, and J. H. S. Torres, Physics of the Solid State 41 (1999) 1959 [Fizika Tverd. Tela (St. Petersburg) 41 (1999) 2135].
[6] Y. Kopelevich, P. Esquinazi, J. H. S. Torres, and S. Moehlecke, J. Low Temp. Phys. 119 (2000) 691.
[7] R. R. da Silva, J. H. S. Torres, and Y. Kopelevich, Phys. Rev. Lett. 87 (2001) 147001.
[8] Yang Hai-Peng, Wen Hai-Hu, Zhao Zhi-Wen, and Li Shi-Liang, Chin. Phys. Lett. 18 (2001) 1648.
[9] S. Moehlecke, Pei-Chun Ho, and M. B. Maple, cond-mat/0204006, to appear in Phil. Mag. Lett. (2002).

[10] J. H. Schön, Ch. Kloc, and B. Batlogg, Nature 408 (2000) 549.
[11] J. H. Schön, Ch. Kloc, and B. Batlogg, Science 293 (2001) 2432.
[12] M. Kociak, A. Yu. Kasumov, S. Gueron, B. Reulet, L. Vaccarini, I. I. Khodos, Yu. B. Gorbatov, V. T. Volkov, and H. Bouchiat, Phys. Rev. Lett. 86 (2001) 2416.
[13] Z. K. Tang, L. Y. Zhang, N. Wang, X. X. Zhang, G. H. Wen, G. D. Li, J. N. Wang, C. T. Chan, and P. Sheng, Science 292 (2001) 2462.
[14] V. I. Tsebro, O. E. Omel´yanovskii, and A. P. Moravskii, JETP. Lett. 70 (1999) 462.
[15] G. M. Zhao and Y. S. Wang, cond-mat/0111268 (to be published in Phil. Mag. B).
[16] J. Nagamatsu, N. Nakagawa, T. Muranaka, Y. Zenitani, and J. Akimitsu, Nature 410 (2001) 63.
[17] S. Mizogami, M. Mizutani, M. Fukuda, and K. Kawabata, Synthetic Metals 41-43 (1991) 3271.
[18] Y. Murakami and H. Suematsu, Pure & Appl. Chem. 68 (1996) 1463.
[19] T. L. Makarova, B. Sundqvist, R. Höhne, P. Esquinazi, Y. Kopelevich, P. Scharff, V. A. Davydov, L. S. Kashevarova, and A. V. Rakhmanina, Nature 413 (2001) 716.
[20] V. E. Antonov, I. O. Bashkin, S. S. Khasanov, A. P. Moravsky, Yu. G. Morozov, Yu. M. Shulga, Yu. A. Ossipyan, and E. G. Ponyatovsky, J. Alloys and Compounds 330-332 (2002) 365.
[21] I. Felner, U. Asaf, Y. Levi, and O. Millo, Phys. Rev. B 55 (1997) 3374.
[22] E. B. Sonin and I. Felner, Phys. Rev. B 57 (1998) 14000.
[23] C. Bernhard, J. L. Tallon, E. Brücher, and R. K. Kremer, Phys. Rev. B 61 (2000) 14960.
[24] S. S. Saxena, P. Agarwal, K. Ahilan, F. M. Grosche, R. K. W. Haselwimmer, M. J. Steiner, E. Pugh, I. R. Walker, S. R. Julian, P. Monthoux, G. G. Lonzarich, A. Huxley, I. Sheikin, D. Braithwaite, & J. Flouquet, Nature 406 (2000) 587.
[25] K. B. Blagoev, J. R. Engelbrecht, and K. S. Bedell, Phys. Rev. Lett. 82 (1999) 133.
[26] N. I. Karchev, K. B. Blagoev, K. S. Bedell, and P. B. Littlewood, Phys. Rev. Lett. 86 (2001) 846.
[27] H. Suhl, Phys. Rev. Lett. 87 (2001) 167007.
[28] A. A. Abrikosov, J. Phys.: Condens. Matter 13 (2001) L943.
[29] A. P. Ramirez, Superconductivity Review 1 (1994) 1.
[30] W. E. Pickett, Solid State Physics 48 (1994) 225.
[31] M. S. Dresselhaus and G. Dresselhaus, Adv. Phys. 30 (1981) 139.
[32] B. T. Kelly, Physics of Graphite, Applied Science, London/New Jersey, 1981.
[33] H. Kempa, Y. Kopelevich, F. Mrowka, A. Setzer, J. H. S. Torres, R. Höhne, and P. Esquinazi, Solid State Commun. 115 (2000) 539.
[34] M. S. Sercheli, Y. Kopelevich, R. R. da Silva, J. H. S. Torres, and C. Rettori, Solid State Commun. 121 (2002) 579.
[35] D. V. Khveshchenko, Phys. Rev. Lett. 87 (2001) 206401; ibid. 87 (2001) 246802.
[36] E. V. Gorbar, V. P. Gusynin, V. A. Miransky, and I. A. Shovkovy, cond-mat/0202422.
[37] J. Gonzales, F. Guinea, and M. A. H. Vozmediano, Phys. Rev. Lett. 77 (1996) 3589.
[38] L. Balents and M. P. A. Fisher, Phys. Rev. B 55 (1997) 11973.
[39] M. P. A. Fisher, Phys. Rev. Lett. 65 (1990) 923.

[40] for a review article see E. Abrahams, S. V. Kravchenko, and M. P. Sarachik, Rev. Mod. Phys. 73 (2001) 251, and references therein.
[41] M. P. A. Fisher, G. Grinstein, and S. M. Girvin, Phys. Rev. Lett. 64 (1990) 587.
[42] M.-C. Cha, M. P. A. Fisher, S. M. Girvin, M. Wallin, and A. P. Young, Phys. Rev. B 44 (1991) 6883.
[43] N. Mason and A. Kapitulnik, Phys. Rev. Lett. 82 (1999) 5341.
[44] D. Das and S. Doniach, Phys. Rev. B 60 (1999) 1261; ibid. 64 (2001) 134511.
[45] D. Dalidovich and P. Phillips, Phys. Rev. Lett. 84 (2000) 737.
[46] D. Dalidovich and P. Phillips, Phys. Rev. B 64 (2001) 052507; ibid. 64 (2001) 184511.
[47] P. Phillips, Phys. Rev. B 64 (2001) 113202.
[48] M. V. Feigel´man and A. I. Larkin, Chem. Phys. 235 (1998) 107.
[49] M. V. Feigel´man, A. I. Larkin, and M. A. Skvortsov, Phys. Rev. Lett. 86 (2001) 1869.
[50] P. Phillips, Y. Wan, I. Martin, S. Knysh, & D. Dalidovich, Nature 395 (1998) 253.
[51] N. Markovic, C. Christiansen, and A. M. Goldman, Phys. Rev. Lett. 81 (1998) 5217.
[52] D. Simonian, S. V. Kravchenko, M. P. Sarachik, and V. M. Pudalov, Phys. Rev. Lett. 79 (1997) 2304.
[53] Y. Hanein, U. Meirav, D. Shahar, C. C. Li, D. C. Tsui, and H. Shtrikman, Phys. Rev. Lett. 80 (1998) 1288.
[54] P. T. Coleridge, R. L. Williams, Y. Feng, and P. Zawadzki, Phys. Rev.B 56 (1997) 12764.
[55] M. S. Dresselhaus and G. Dresselhaus, Adv. Phys. **30**, 139 (1981).
[56] V. M. Pudalov, G. Brunthaler, A. Prinz, and G. Bauer, Physica B 251 (1998) 697.
[57] X. P. A. Gao, A. P. Mills, Jr., A. P. Ramirez, L. N. Pfeiffer, and K. W. West, Phys. Rev. Lett. 88 (2002) 166803.
[58] H. S. J. van der Zant, W. J. Elion, L. J. Geerligs, and J. E. Mooij, Phys. Rev. B 54 (1996) 10081.
[59] F. C. Zhang and T. M. Rice, cond-mat/9708050.
[60] T. M. Rice, Nature 389 (1997) 916.
[61] Y. H. Chen, F. Wilczek, E. Witten, and B. I. Halperin, Int. J. Mod. Phys. 3 (1989) 1001.
[62] A. G. Aronov and A. D. Mirlin, Phys. Lett. A 152 (1991) 371.
[63] L. Pauling, Nature of the Chemical Bond (Cornell University Press, New York, 1960).
[64] G. Baskaran and S. A. Jafari, cond-mat/0110022.
[65] V. Kalmeyer and R. B. Laughlin, Phys. Rev. Lett. 59 (1987) 2095.
[66] R. B. Laughlin, Phys. Rev. Lett. 60 (1988) 2677.
[67] D. H. Lee and M. P. A. Fisher, Phys. Rev. Lett. 63 (1989) 903.
[68] Y. Ren and F. C. Zhang, Phys. Rev. B 49 (1994) 1532.
[69] B. I. Halperin and T. M. Rice, Rev. Mod. Phys. 40 (1968) 755.
[70] A. A. Abrikosov, J. Less-Common Metals 62 (1978) 451.

[71] J. González, F. Guinea, and M. A. H. Vozmediano, Phys. Rev. B 63 (2001) 134421.
[72] J. A. Woollam, Phys. Rev. Lett. 25 (1970) 810.
[73] Y. Iye, P. M. Tedrow, G. Timp, M. Shayegan, M. S. Dresselhaus, G. Dresselhaus, A. Furukawa, and S. Tanuma, Phys. Rev. B 25 (1982) 5478.
[74] G. Timp, P. D. Dresselhaus, T. C. Chieu, G. Dresselhaus, and Y. Iye, Phys. Rev. B 28 (1983) 7393.
[75] Y. Iye, L. E. McNeil, and G. Dresselhaus, Phys. Rev. B 30 (1984) 7009.
[76] Y. Kopelevich, V. V. Makarov, and L. M. Sapozhnikova, Fiz. Tv. Tela 26 (1984) 2651; Sov. Phys. Solid State 26 (1984)1607.
[77] M. Büttiker, Phys. Rev. Lett. 57 (1986) 1761.
[78] A. D. Benoit, S. Washburn, C. P. Umbach, R. B. Laibowitz, and R. A. Webb, Phys. Rev. Lett. 57 (1986) 1765.
[79] G. Timp, H. U. Baranger, P. deVegvar, J. E. Cunningham, R. E. Howard, R. Behringer, and P. M. Mankiewich, Phys. Rev. Lett. 60 (1988) 2081.
[80] P. G. N. de Vegvar and T. A. Fulton, Phys. Rev. Lett. 71 (1993) 3537.
[81] H. P. Wei, D. C. Tsui, M. A. Paalanen, and A. M. M. Pruisken, Phys. Rev. Lett. 61 (1988) 1294.
[82] H. W. Jiang, C. E. Johnson, K. L. Wang, and S. T. Hannahs, Phys. Rev. Lett. 71 (1993) 1439.
[83] T. Wang, K. P. Clark, G. F. Spencer, A. M. Mack, and W. P. Kirk, Phys. Rev. Lett. 72 (1994) 709.
[84] D. Shahar, D. C. Tsui, M. Shayegan, R. N. Bhatt, and J. E. Cunningham, Phys. Rev. Lett. 74 (1995) 4511.
[85] S. H. Song, D. Shahar, D. C. Tsui, Y. H. Xie, and D. Monroe, Phys. Rev. Lett. 78 (1997) 2200.
[86] C. H. Lee, Y. H. Chang, Y. W. Suen, and H. H. Lin, Phys. Rev. B 56 (1997) 15238.
[88] M. Hilke, D. Shahar, S. H. Song, D. C. Tsui, Y. H. Xie, and D. Monroe, Phys. Rev. B 56 (1997) R15545.
[89] C. H. Lee, Y. H. Chang, Y. W. Suen, and H. H. Lin, Phys. Rev. B 58 (1998) 10629.
[90] M. Hilke, D. Shahar, S. H. Song, D. C. Tsui, and Y. H. Xie, Phys. Rev. B 62 (2000) 6940.
[91] C. F. Huang, Y. H. Chang, C. H. Lee, H. T. Chou, H. D. Yeh, C. T. Liang, Y. F. Chen, H. H. Lin, H. H. Cheng, and G. J. Hwang, Phys. Rev. B 65 (2001) 045303.
[92] S. T. Hannahs, J. S. Brooks, W. Kang, L. Y. Chiang, and P. M. Chaikin, Phys. Rev. Lett. 63 (1989) 1988.
[93] S. Hill, S. Uji, M. Takashita, C. Terakura, T. Terashima, H. Aoki, J. S. Brooks, Z. Fisk, and J. Sarrao, Phys. Rev. B 58 (1998) 10778.
[94] J. T. Chalker and A. Dohmen, Phys. Rev. Lett. 75 (1995) 4496.
[95] L. Balents and M. P. A. Fisher, Phys. Rev. Lett. 76 (1996) 2782.
[96] J. D. Naud, L. P. Pryadko, and S. L. Sondhi, Phys. Rev. Lett. 85 (2000) 5408.
[97] Z. Tesanovic, M. Rasolt, and L. Xing, Phys. Rev. Lett. 63 (1989) 2425; Phys. Rev. B 43 (1991) 288.

[98] H. Akera, A. H. MacDonald, S. M. Girvin, and M. R. Norman, Phys. Rev. Lett. 67 (1991) 2375.
[99] A. H. MacDonald, H. Akera, and M. R. Norman, Phys. Rev. B 45 (1992) 10147.
[100] M. R. Norman, H. Akera, and A. H. MacDonald, Physica C 196 (1992) 43.
[101] A. H. MacDonald, H. Akera, and M. R. Norman, Aust. J. Phys. 46 (1993) 333.
[102] H. Akera, A. H. MacDonald, and M. R. Norman, Physica B 184 (1993) 337.
[103] H. Akera, A. H. MacDonald, and D. Yoshioka, Physica B 201 (1994) 255.
[104] M. Rasolt and Z. Tesanovic, Rev. Mod. Phys. 64 (1992) 709.
[105] for a review article see E. A. Pashitskii, Low Temp. Phys. 25 (1999) 690.
[106] V. W. Scarola, K. Park, and J. K. Jain, Nature 406 (2000) 863.
[107] N. Read and D. Green, Phys. Rev. B 61 (2000) 10267.
[108] S. Mo and A. Sudbo, cond-mat/0111279.
[109] M. M. Maska, cond-mat/0202247.
[110] H. Kempa, P. Esquinazi, and Y. Kopelevich, Phys. Rev. B (2002), in press; cond-mat/0204166.
[111] Y. Iye and G. Dresselhaus, Phys. Rev. Lett. 54 (1985) 1182.
[112] H. Yaguchi and J. Singleton, Phys. Rev. Lett. 81 (1998) 5193.
[113] Y. Shimamoto, N. Miura, and H. Nojiri, J. Phys.: Cond. Matter 10 (1998) 11289.
[114] Y. Takada and H. Goto, J. Phys.: Cond. Matter 10 (1998) 11315.
[115] S. Uji, J. S. Brooks, and Y. Iye, Physica B 246-247 (1998) 299.
[116] H. Yaguchi, T. Takamasu, Y. Iye, and N. Miura, J. Phys. Soc. Jap. 68 (1999) 181.
[117] H. Yaguchi, J. Singleton, and T. Iwata, Physica B 298 (2001) 546.
[118] M. A. H. Vozmediano, M. P. López Sancho, and F. Guinea, cond-mat/0110418.
[119] P. Moses and R. H. McKenzie, Phys. Rev. B 60 (1999) 7998.
[120] H. Kempa, P. Esquinazi and Y. Kopelevich, to be published.
[121] C. Ayache, Ph.D. Thesis, Grenoble (1978, unpublished).
[122] J. A. Woollam, Phys. Rev. B 3 (1971) 1148.
[123] R. Ocaña and P. Esquinazi, to be published.
[124] E. J. Ferrer, V. P. Gusynin and V. de la Incera, cond-mat/0203217.
[125] P. Esquinazi, A. Setzer, R. Höhne, C. Semmelhack, Y. Kopelevich, D. Spemann, T. Butz, B. Kohlstrunk, and M. Loesche, cond-mat/0203153.
[126] G. Deutscher and P. G. de Gennes, in Superconductivity, edited by R. Parks (Marcel Dekker, New York, 1969), Vol. 2.
[127] G. Deutscher and A. Kapitulnik, Physica A 168 (1990) 338.
[128] A. L. Fauchère and G. Blatter, Phys. Rev. B 56 (1997) 14102.
[129] C. Ebner and D. Stroud, Phys. Rev. B 31 (1985) 165.
[130] M. Fujita, K. Wakabayashi, K. Nakada, and K. Kushakabe, J. Phys. Soc. Jpn. 65 (1996) 1920.
[131] M. Fujita, M. Igami, and K. Nakada, J. Phys. Soc. Jpn. 66 (1997) 1864.
[132] K. Nakada, M. Fujita, G. Dresselhaus G, and M.S. Dresselhaus, Phys. Rev. B 54 (1996) 17954.
[133] P. L. Giunta and S. P. Kelty, J. Chem. Phys. 114 (2001) 1807.
[134] K. Wakabayashi, M. Fujita, H. Ajiki, and M. Sigrist, Physica B 280 (2000) 388.
[135] Y. Shibayama, H. Sato, T. Enoki, and M. Endo, Phys. Rev. Lett. 84 (2000) 1744.
[136] K. J. Harigaya, J. Phys.: Condens. Matter 13 (2001) 1295.

[137] F. D. M. Haldane, Phys. Rev. Lett. 61 (1988) 2015.
[138] A. Kapitulnik, N. Mason, S. A. Kivelson, and S. Chakravarty, Phys. Rev. B 63 (2001) 125322.
[139] E. Wigner, Trans. Faraday Soc. 34 (1938) 678.